\documentclass[a4paper,11pt]{article}
\pdfoutput=1 % if your are submitting a pdflatex (i.e. if you have
\usepackage{jheppub} % for details on the use of the package, please
\usepackage[T1]{fontenc} % if needed

\usepackage{comment}
\usepackage{mathrsfs}
\usepackage{mathtools}

\usepackage{graphicx}
\usepackage{bm}
\usepackage{amssymb}
\usepackage{amsmath}
\usepackage{cancel}
\usepackage{hyperref}
\usepackage{color}
\usepackage{dsfont}
\usepackage{slashed}
\usepackage{enumerate}

\usepackage{soul}

\def\shrinkage{2.1mu}
\def\vecsign{\mathchar"017E}
\def\dvecsign{\smash{\stackon[-1.95pt]{\mkern-\shrinkage\vecsign}{\rotatebox{180}{$\mkern-\shrinkage\vecsign$}}}}
\def\dvec#1{\def\useanchorwidth{T}\stackon[-4.2pt]{#1}{\,\dvecsign}}
\usepackage{stackengine}
\stackMath
\usepackage{graphicx}

\newcommand{\Beq}{\begin{equation}\begin{aligned}}
\newcommand{\Eeq}{\end{aligned}\end{equation}}

\def\({\left(}
\def\){\right)}
\def\[{\left[}
\def\]{\right]}

\def\nn{\nonumber}
\def\({\left(}
\def\){\right)}
\def\[{\left[}
\def\]{\right]}

\title{Axion anomalies}

\author{Peter Adshead}
\author{and Kaloian D. Lozanov}

\affiliation{Illinois Center for Advanced Studies of the Universe \& Department of Physics, University of Illinois at Urbana-Champaign, Urbana, IL 61801, USA.}

\emailAdd{adshead@illinois.edu}
\emailAdd{klozanov@illinois.edu}

\abstract{
    We study fermions derivatively coupled to axion-like or pseudoscalar fields, and show that  the axial vector current of the fermions is not conserved in the limit where the fermion is massless. This apparent violation of the classical chiral symmetry is due to the background axion field.  We compute the contributions to this anomalous Ward identity due to the pseudoscalar field alone, which arise in Minkowski space, as well as the effects due to an interaction with an external gravitational field. For the case of massless fermions, these interactions induce terms in the axion effective action that can be removed by the addition of local counterterms. 
 We demonstrate that these counterterms are generated by the transformation of the path integral measure when transforming the theory from a form where the chiral symmetry is manifest to one where the symmetry is only apparent after using the classical equations of motion.
We work perturbatively in Minkowski space and include the effects of interactions with a linearized gravitational field. Using the heat kernel method, we study the transformation properties of the path integral measure, and include the effects of non-linear gravity as well as interactions with gauge fields. Finally, we verify our relation by considering derivatively coupled fermions during pseudoscalar-driven inflation and computing the divergence of the axial current in de Sitter spacetime.
}

\begin{document}
\maketitle
\flushbottom

%-----------------------------
%-----Introduction------------
%-----------------------------

\section{Introduction}

Anomalies in quantum field theories have a long history.  Most famously, Adler, Jackiw and Bell showed that the axial current is anomalous in quantum electrodynamics due to the triangle diagram \cite{Adler:1969gk,Bell:1969ts}, and that this process gives the correct observed lifetime for the decay of the neutral pion  to two photons, $\pi^0 \to 2\gamma$.  Adler and Bardeen later showed that the perturbative result from the triangle diagram was one-loop exact \cite{Adler:1969er}. More generally, Bardeen  showed that the axial current generally has anomalous Ward identities due to couplings to scalar, pseudo-scalar, vector and axial-vector fields \cite{Bardeen:1969md}. He showed that only those anomalies that have abnormal parity---those corresponding to the vector fields---are essential and cannot be removed via counterterms without spoiling the gauge symmetry.  Fujikawa later demonstrated that anomalies arise due to the transformation properties of the measure of the path integral \cite{Fujikawa:1979ay, Fujikawa:1980eg}. In particular, Fujikawa elegantly showed that anomalies arise when symmetries of the classical action are not respected by the path integral measure. The anomaly has been further verified by solving the fermion equation of motion in an Abelian \cite{Nielsen:1983rb} and a non-Abelian \cite{Domcke:2018gfr} gauge field background. Gravitational corrections to the partially conserved axial current (PCAC) were considered {by Kimura in Ref. \cite{10.1143/PTP.42.1191}, and} by Salam and Delbourgo in Refs.\ \cite{Delbourgo:1972xb, Delbourgo:1972en}, while Witten and Alvarez-Gaume studied gravitational anomalies more generally in Ref.\ \cite{Alvarez-Gaume:1983ihn}. The gravitational anomaly in the standard model lepton current has been considered as a possible source of the matter-antimatter asymmetry \cite{Alexander:2004us, Adshead:2017znw}. Violation of the $B+L$ number also occurs through the chiral non-Abelian anomaly during electroweak sphaleron processes \cite{Klinkhamer:1984di}, and is a key component of many mechanisms that generate the baryon asymmetry in the early Universe \cite{Harvey:1990qw}.

In this paper, motivated by theories of the early Universe containing axion fields with classical vacuum expectation values interacting with fermions  \cite{Adshead:2015kza, Adshead:2015jza, Adshead:2018oaa, Adshead:2019aac, Wang:2019gbi, Roberts:2021plm,Domcke:2018eki,Domcke:2019mnd,Domcke:2019qmm,Domcke:2020kcp,Domcke:2021fee,Domcke:2021yuz,Mirzagholi:2019jeb,Maleknejad:2019hdr,Maleknejad:2020yys,Maleknejad:2020pec,Agrawal:2018mkd}, we revisit anomalous Ward identities due to axion or pseudoscalar interactions.

We consider the theory of a single generation of Dirac fermions interacting with a pseudoscalar or axion-like field described by the action
\begin{align}\label{eqn:fermaxaction}
S \equiv S_\phi + S_f = \int d^4 x\sqrt{-g}\[  \frac{1}{2}(\partial\phi)^2 - V(\phi) + i \bar\psi \gamma^{\mu}D_{\mu} \psi-m \bar{\psi}e^{2i\gamma_5\phi/f}\psi\].
\end{align}
We assume the axion-like field has a potential due to some unspecified symmetry breaking.  Here, $D_{\mu} $ is the covariant derivative, $D_{\mu} = \partial_\mu +\omega_{ab}\Sigma^{ab}+i A_\mu$, where $A_\mu$ is a gauge field,  $\omega_{ab}$ is the spin connection, and $\Sigma^{ab}$ are the generators of the Lorentz group.

Under axial or chiral rotations of the fermion
\begin{align}\label{eqn:axialrot}
\psi \to \chi =  e^{i\gamma_5  \alpha(x)}\psi ,
\end{align}
the (classical) fermion action transforms to
\begin{align}\label{eqn:fermaxaction2}
S_f =  & \int d^4 x \sqrt{-g}\Big[ i\bar{\chi}\gamma^{\mu}D_{\mu} \chi-m \bar{\chi}\chi+\frac{\partial_{\mu}\phi }{f}\bar{\chi}\gamma^{\mu}\gamma_5\chi\Big]\,,
\end{align}
provided we set $\alpha=\phi/f$. 

If the fermion is massless $m = 0$, notice that the interaction between the fermion and the axion in eq. \eqref{eqn:fermaxaction} is removed. This observation implies that the massless limit of the theory in eq.\ \eqref{eqn:fermaxaction2} should not exhibit any effects of the axion-fermion interaction. It is tempting to show this by integrating the interaction in eq.\ \eqref{eqn:fermaxaction2} by parts, and using the classical equations of motion
\begin{align}\label{eqn:classicalaxial}
\partial_\mu j^\mu_5  \equiv \partial_\mu (\bar{\chi}\gamma^{\mu}\gamma_5\chi)  = 2 i m\bar{\chi}\gamma_5\chi\,,
\end{align}
to make this manifest. In the limit $m \to 0$, eq.\ \eqref{eqn:classicalaxial} is the conservation law associated with the axial transformation in eq.\ \eqref{eqn:axialrot} which is a symmetry of the classical action in this limit.  

In the quantum theory,  the situation is more complicated. In the presence of a gauge field, or a gravitational field, the axial current is no longer conserved due to the anomaly. The divergence of the axial current due to the Adler-Jackiw-Bell \cite{Adler:1969gk, Bell:1969ts} and gravitational anomalies \cite{Delbourgo:1972xb, Alvarez-Gaume:1983ihn}, reads
\Beq
\label{eq:AxionAnomalyCurrent1}
\langle \nabla_\mu j^{\mu}_5(x)\rangle=-\frac{1}{16\pi^2}\epsilon^{\mu\nu\alpha\beta}F_{\mu\nu}F_{\alpha\beta}+\frac{1}{384\pi^2}\epsilon^{\mu\nu\alpha\beta}R^{\rho\sigma}{}_{\mu\nu}R_{\rho\sigma\alpha\beta}\,.
\Eeq
In the massless limit, this anomaly means that fermion mediated axion interactions with gauge fields and gravitons are non-vanishing when computed using the theory in eq. \eqref{eqn:fermaxaction2}, in contrast to the expectation from the theory in eq. \eqref{eqn:fermaxaction}. However, the theories in eqs.\ \eqref{eqn:fermaxaction} and \eqref{eqn:fermaxaction2}
 are still equivalent once one takes into account the Jacobian factor in the path integral measure for the transformation that takes eq. \eqref{eqn:fermaxaction} into eq. \eqref{eqn:fermaxaction2}.  In particular, a rotation by $\alpha=\phi/f$ leads to the additional contributions to the action
 \begin{align}
\Delta \mathcal{L}_{\rm jac} = \frac{\phi}{16\pi^2f}\epsilon^{\mu\nu\alpha\beta}F_{\mu\nu}F_{\alpha\beta}-\frac{\phi}{384\pi^2f}\epsilon^{\mu\nu\alpha\beta}R^{\rho\sigma}{}_{\mu\nu}R_{\rho\sigma\alpha\beta}.
\end{align}
These additional interactions precisely cancel the effects of the anomaly in the axial current (for a recent discussion, see \cite{Quevillon:2019zrd}).

The purpose of this paper is to demonstrate that in the presence of the axion field, the conservation law in eq.\ \eqref{eqn:classicalaxial} contains additional anomalous terms in the quantum mechanical theory.  These ``naively anomalous'' additional terms were pointed out by Bardeen in Ref.\ \cite{Bardeen:1969md}, however, we also demonstrate that further anomalies involving the axion appear in the theory when it is coupled to gravity. In particular, we demonstrate that, as well as the usual anomalies associated with gauge fields and gravity (in eq.\ \eqref{eq:AxionAnomalyCurrent1}),  eq.\ \eqref{eqn:classicalaxial} receives contributions from the axion, as well as mixed contributions from the axion and  the gravitational field. The final renormalized result in the quantum theory for the theory in eq.\ \eqref{eqn:fermaxaction2} when $m\rightarrow0$ is (see eq.\ \eqref{eqn:heatanomy})
\Beq
\label{eq:AxionAnomalyCurrent}
\langle \nabla_\mu j^{\mu}_5(x)\rangle=-&\frac{\Box^2 \phi}{12\pi^2f}-\frac{\nabla_\mu\(G^{\mu\nu}\partial_\nu\phi\)}{12\pi^2f}+\frac{1}{3\pi^2}\nabla_\mu \left(\frac{\partial^\mu\phi}{f}\frac{\partial_\nu\phi}{f}\frac{\partial^\nu\phi}{f}\right)\\  -&\frac{1}{16\pi^2}\epsilon^{\mu\nu\alpha\beta}F_{\mu\nu}F_{\alpha\beta}+\frac{1}{384\pi^2}\epsilon^{\mu\nu\alpha\beta}R^{\rho\sigma}{}_{\mu\nu}R_{\rho\sigma\alpha\beta}\,,
\Eeq
%where we have also included the possibility the fermion is charged under a gauge field with field strength $F_{\mu\nu}$. 
In this expression $G^{\mu\nu}$ is the Einstein tensor,  $R_{\rho\sigma\alpha\beta}$ is the Riemann tensor,  $\Box = \nabla^\alpha \nabla_\alpha$, and $\nabla_\alpha$ is the covariant derivative. The terms on the right-hand side involving the axion are all total derivatives, and are purely local functions. They are a fermion one-loop effect and induce 2-, 3- and 4-point, $\sim(\Box \phi/f)^2$, $\sim G^{\mu\nu}\partial_\mu\phi\partial_\nu\phi/f^2$, $\sim(\partial \phi/f)^4$, interaction terms in the effective curved spacetime axion action. 

The violation of the conservation law in eq.~\eqref{eq:AxionAnomalyCurrent} in the quantum theory leads to the non-vanishing of fermion-mediated axion correlation functions in the massless limit. This is in contradiction with our expectations from the original theory, eq. \eqref{eqn:fermaxaction}, where the fermions manifestly decouple from the axion when $m\rightarrow0$.

In this paper, we demonstrate  that, as in the case with the gauge and gravitational fields, the reconciliation of the two results comes from the fact that the path integral measure is not invariant under the chiral rotation in eq.\ \eqref{eqn:axialrot}. The Jacobian factor from this chiral rotation yields additional terms in eq.\ \eqref{eqn:fermaxaction2}, and reads
\begin{align}\nn
\Delta \mathcal{L}_{\rm jac} = &\frac{1}{12\pi^2}\left[\frac{1}{2}G^{\mu\nu}\frac{\partial_\mu\phi}{f}\frac{\partial_\nu\phi}{f}-\left(\frac{\partial\phi}{f}\right)^4-\frac{1}{2}\left(\frac{\Box \phi}{f}\right)^2\right]\\
&-\frac{\phi}{384\pi^2 f}\epsilon^{\mu\nu\alpha\beta}R^{\rho\sigma}{}_{\mu\nu}R_{\rho\sigma\alpha\beta}+\frac{\phi}{16\pi^2 f}\epsilon^{\mu\nu\alpha\beta}F_{\mu\nu}F_{\alpha\beta} .
\end{align}
On the one hand, when $m\rightarrow0$, these Jacobian terms exactly cancel the interactions in the axion effective action induced by the anomaly.  On the other hand, when $m \to \infty$, $\Delta \mathcal{L}_{\rm jac}$ are the terms induced in the effective action  from integrating out the fermion. The actions in eqs.\  \eqref{eqn:fermaxaction} and \eqref{eqn:fermaxaction2} therefore describe the same physical theory, provided the latter is supplemented by the Jacobian terms coming from the non-invariance of the fermion path integral measure under chiral rotations in the axion background. 

The rest of this paper is organized as follows. In section \ref{sec:minkowski}, we compute the perturbative contributions to the divergence of the axial vector current which lead to violations of eq.\ \eqref{eqn:classicalaxial}, and in section \ref{sec:pertgrav} we compute the perturbative  gravitational corrections to the results from  section \ref{sec:minkowski}. In section~\ref{sec:desitter}, we compute the divergence of the axial current for a Dirac fermion coupled to a slowly rolling pseudoscalar in a de Sitter background. We regularize the current using Pauli-Villars fields, as well as adiabatic subtraction, and verify the anomaly equation, including the gravitational contributions in both cases.  In section \ref{sec:pathint}, we show that the measure for the path integral is not invariant under chiral rotations of the fermion in the presence of the axion field. We compute the anomaly function that results from this rotation in two ways. The method in section \ref{sec:pathmom} expands the anomaly function in momentum space, and we reproduce the results from the perturbative calculations in Minkowski space from section~\ref{sec:minkowski}. The second method in section~\ref{sec:pathheat}, uses the heat kernel approach and includes the effects of background gravitational and gauge fields to reproduce, and go beyond the results from section~\ref{sec:pertgrav}. In section \ref{sec:Counterterms} we show that the correlation functions for the axion field are the same in the $\psi$ and $\chi$ formulations, provided we account for the Jacobian terms associated with the chiral transformation relating the two. Finally, we conclude in section \ref{sec:conclusions}. Appendix \ref{app:adiabatic} outlines the adiabatic solutions used to regularize the de Sitter current in section \ref{sec:desitter}.

%-------------------------------------------------------------
%-----Anomalous Ward Identities in Minkowski space------------
%-------------------------------------------------------------

\section{Anomalous Ward identities in Minkowski space}\label{sec:minkowski}

\begin{figure}
\centering
\includegraphics[scale = 0.8]{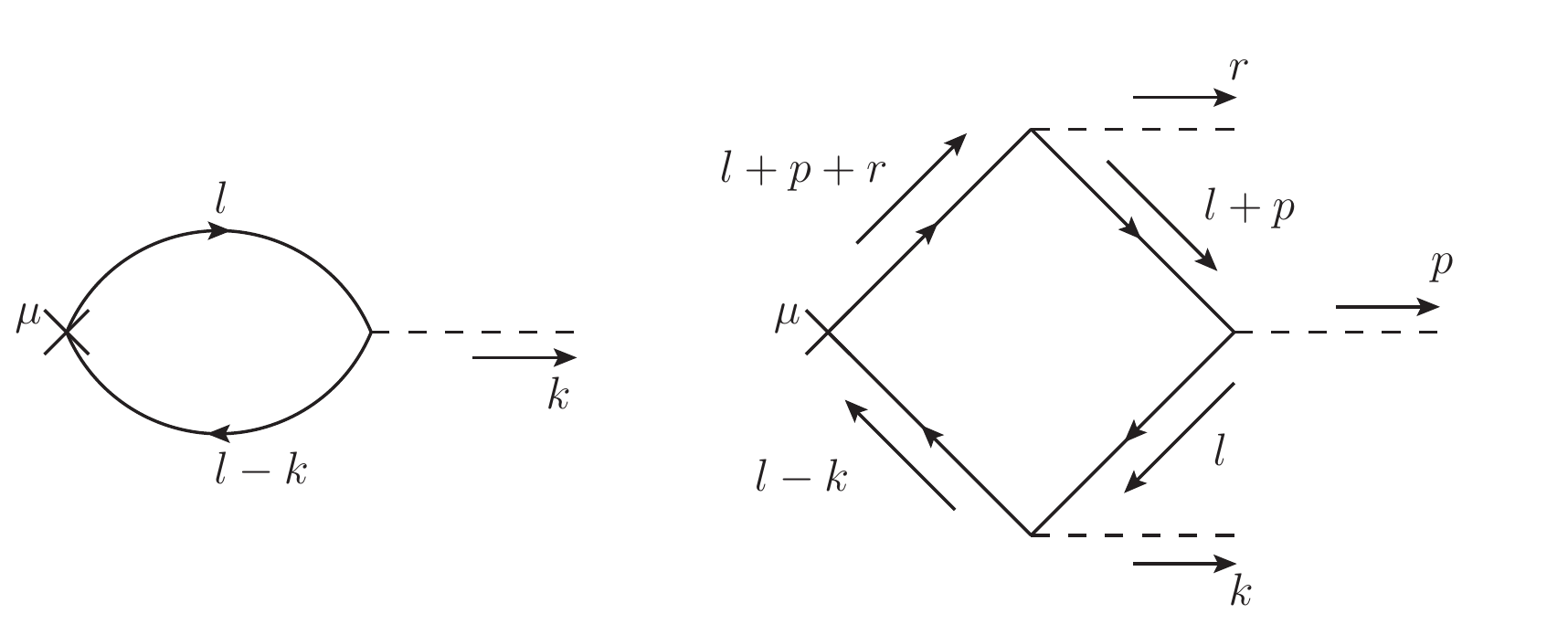}
\caption{Contributions to the divergence of the axial vector current due to emission of axion-like, or pseudoscalar particles.}\label{fig:minkowskidiags}
\end{figure}

We begin by considering perturbative processes in the theory in eq.\ \eqref{eqn:fermaxaction2} that lead to the quantum mechanical  violation of the Ward identity that follows from eq.\ \eqref{eqn:classicalaxial}. We ignore the gauge field in this section, focussing on pure axionic effects. As in the standard textbook treatment, we study the amplitude for the divergence of the axial-vector current to create external axions. The contributions to this process arise from matrix elements of two and four axial-vector vertices, to which are attached external axion lines. The processes are displayed in figure \ref{fig:minkowskidiags}. 

\begin{figure}
\centering
\includegraphics[scale = 0.8]{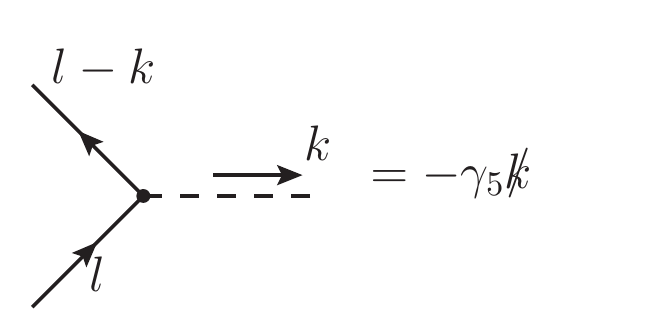}
\caption{Feynman rule for pseudoscalar-fermion interaction vertex.}\label{fig:minkowskirules}
\end{figure}

To evaluate the diagrams in figure \ref{fig:minkowskidiags}, we use the Feynman rules and notational conventions from Peskin and Schroeder \cite{Peskin:1995ev}. We need to supplement with the Feynman rule for the axion vertex, this is shown in figure \ref{fig:minkowskirules}. Every external axion line comes with a factor of $\phi/f$. We now compute each diagram in figure \ref{fig:minkowskidiags}, however, we omit many details which are straightforward to reproduce using standard techniques.

%-------------------------------------------------------------
%-----1 axion processes---------------------------------------
%-------------------------------------------------------------

\subsection{One-axion processes}

We begin with the first diagram in figure \ref{fig:minkowskidiags}, which generates a matrix element for the vacuum to one axion processes 
\begin{align}
\int d^4 x e^{- i q\cdot x}\langle k | j^{\mu}_5(x)|0\rangle  = (2\pi)^4\delta^{(4)}(k-q) \mathcal{M}_1^{\mu}(k).
\end{align}
The diagram corresponds to the integral
\begin{align}\label{eq:oneaxion}
\mathcal{M}_1^\mu(k) = & - \int \frac{d^4 l}{(2\pi)^4}{\rm tr}\[\gamma^\mu \gamma_5 \frac{i((\slashed{l}-\slashed{k})+m)}{(l - k)^2-m^2}(-\gamma_5\slashed{k})\frac{i(\slashed{l}+m)}{l^2-m^2}\].
\end{align}
This integral is divergent and needs to be regularized. We regulate the integral by introducing a set of Pauli-Villars  fields with masses $M_i$ and compute
\begin{align}
\mathcal{M}^\mu_{1, \rm reg}(k) =\mathcal{M}_1^\mu(k) + \sum_i c_i \mathcal{M}_{1,i}^\mu(k),
\end{align}
where $\mathcal{M}_{1,i}^\mu(k)$ is the amplitude in eq.\ \eqref{eq:oneaxion} with $m \to M_i$. To remove the divergences, we impose the conditions
\begin{align}
\sum_i c_i = -1, \quad \sum_i c_i M_i^2 = -m^2 .
\end{align}
We are ultimately interested in the divergence of the axial current, which is equivalent to dotting $\mathcal{M}_{1,\rm reg}^\mu(k)$ with $iq_{\mu}$. Using the identity 
\begin{align}
\slashed{q} \gamma_5 = (\slashed{l}-m)\gamma_5 + \gamma_5 (\slashed{l}-\slashed{k}-m) + 2\gamma_5 m,
\end{align}
and relabeling the loop momenta we can eliminate one $\gamma$-matrix from the matrix element in eq.\ \eqref{eq:oneaxion} to obtain
\begin{align}\label{eq:regoneax}
iq_\mu \mathcal{M}_{1, \rm reg}^\mu(k) =  -i\int \frac{d^4 l}{(2\pi)^4}{\rm tr}\Bigg[ 2m \gamma_5 \frac{(\slashed{l}-\slashed{k}+m)}{(l - k)^2-m^2}\gamma_5\slashed{k}\frac{\slashed{l}+m}{(l^2-m^2)}+\sum_i c_i (m\to M_i)\Bigg].
\end{align}
Evaluating the expression in eq.\ \eqref{eq:regoneax} by the usual methods,  and taking the limit $m \to 0$ and $M_i \to \infty$, we get
\begin{align}
iq_\mu \mathcal{M}_{1, \rm reg}^\mu(k)
= -k^2\frac{1}{2\pi^2}\[ \sum_i c_i M_i^2 \log\[M_i^2\] +\frac{1}{6}k^2 \],
\end{align}
where this contribution arises purely from the regulator fields.  This amplitude implies
\begin{align}
\langle k| \partial_\mu j^{\mu}_5(0)|0\rangle =   \frac{1}{2\pi^2} \langle k | \[  \sum_i c_i M_i^2 \log\[M_i^2\] \frac{\Box\phi(0)}{f} -\frac{1}{6}\frac{1}{f} \Box^2 \phi(0) \]|0\rangle  .
\end{align}
Note that this result is divergent as $M_i \to \infty$. However, the divergence can be absorbed without introducing a new counterterm by renormalizing of the axion kinetic term, leaving us with a finite new term proportional to $\Box^2 \phi$.

%-------------------------------------------------------------
%-----3 axion processes---------------------------------------
%-------------------------------------------------------------

\subsection{Three-axion processes}

We next consider the contribution from the right diagram in figure \ref{fig:minkowskidiags}. This graph leads to an anomalous contribution to the divergence of the axial current given by
\begin{align}
\int d^4 x e^{- i q\cdot x}\langle k,p,r | j^{\mu}_5(x)|0\rangle  = (2\pi)^4\delta^{(4)}(k+p+r-q) \mathcal{M}_3^\mu(k, p, r)
\end{align}
where
\begin{align}\nn
 \mathcal{M}_3^\mu = 
& \int \frac{d^4 l}{(2\pi)^4}{\rm tr}\Bigg[\gamma^\mu \gamma_5 \frac{i(\slashed{l}-\slashed{k} + m)}{(l - k)^2-m^2}\gamma_5\slashed{k} \frac{i(\slashed{l}+m)}{(l^2 - m^2)}\gamma_5\slashed{p}\frac{i(\slashed{l}+\slashed{p}+m)}{(l+p)^2-m^2}\gamma_5\slashed{r}\frac{i(\slashed{l}+\slashed{p}+\slashed{r}+m)}{(l+p+r)^2-m^2}\Bigg] \\ & + {\rm perms}
\end{align}
where `perms' indicates diagrams obtained by exchanging pairs of external momenta. There are 6 diagrams in total. As above, we regularize by introducing Pauli-Villars fields, 
\begin{align}
\mathcal{M}_{3, \rm reg}^\mu(k,p,r) =\mathcal{M}_3^\mu(k,p,r) + \sum_i c_i \mathcal{M}_{3,i}^\mu(k,p,r).
\end{align}
We again contract $\mathcal{M}_{3, \rm reg}^\mu(k,p,r)$ with $iq_{\mu}$, and write
\begin{align}
\slashed{q} \gamma_5 =(\slashed{k}+{\slashed{p}}+\slashed{r})\gamma_5 =  (\slashed{l}+{\slashed{p}}+\slashed{r} - m)\gamma_5 + \gamma_5 (\slashed{l}-\slashed{k} - m) + 2\gamma_5m.
\end{align}
Inserting this, we find (after relabelling the loop momenta, and summing diagrams to cancel terms)
\begin{align}\nn
&  i q_\mu \mathcal{M}_{3, \rm reg}^\mu(k,p,r)\\\nn
&= i  \int \frac{d^4 l}{(2\pi)^4}{\rm tr}\Bigg[ 2m\gamma_5 \frac{(\slashed{l}-\slashed{k}+m)}{(l - k)^2-m^2}\gamma_5 \slashed{k}\frac{\slashed{l}+m}{(l^2-m^2)}\gamma_5\slashed{p} \frac{(\slashed{l}+\slashed{p}+m)}{(l+p)^2-m^2}\gamma_5\slashed{r} \frac{(\slashed{l}+\slashed{p}+\slashed{r}+m)}{(l+p+r)^2-m^2}\Bigg]\\ &  + \sum_i c_i(m \to M_i)+{\rm perms}.
\end{align}
This expression can be evaluated using the standard techniques to obtain, in the limit $m\to 0$, $M_i \to \infty$
\begin{align}
i q_\mu \mathcal{M}_{3, \rm reg}^\mu(k,p,r) 
= \frac{1}{2\pi^2}\frac{4}{3}  {k}_\alpha{p}_\beta {r}_\nu q_\mu (\eta^{\alpha\nu}\eta^{\beta\mu} + \eta^{\alpha\beta}\eta^{\nu\mu} + \eta^{\alpha\mu}\eta^{\beta\nu}),
\end{align}
which implies
\begin{align}
\langle k,p,r | \partial_\mu j^{\mu}_5(0)|0\rangle =  \frac{1}{2\pi^2}\frac{2}{3}  \langle k ,p,r | \partial^\mu\(\frac{ \partial_\mu\phi(0)}{f} \( \frac{ \partial\phi(0)}{f}\)^2\) |0\rangle  .
\end{align}
Note that an extra factor of $1/3!$ comes from the number of ways of attaching the external lines.

%-------------------------------------------------------------
%-----Total perturbative Minkowski results--------------------
%-------------------------------------------------------------

\subsection{Total perturbative Minkowski result}

Taken together, the results of the previous section imply that, in the presence of an axion, and absence of gravity or gauge fields, the divergence of the axial current should read
\begin{align}\label{eqn:minkanom}
 \partial_\mu j^{\mu}_5 = &  \frac{1}{2\pi^2} \Bigg[ \sum_i c_i M_i^2 \log\[M_i^2\]  \frac{\Box\phi}{f} -\frac{1}{6}\frac{1}{f} \Box^2 \phi + \frac{2}{3} \partial^\mu\(\frac{ \partial_\mu\phi}{f} \( \frac{ \partial\phi}{f}\)^2\) \Bigg].
\end{align}
As noted above, the divergent term can be absorbed into the wavefunction renormalization of the axion field, while the contributions of the second and third terms to physical processes can be removed by the addition of the counterterms
\begin{align}
\mathcal{L}_{\rm c.t.} = -\frac{1}{24\pi^2}\left(\frac{\Box \phi}{f}\right)^2 -\frac{1}{12\pi^2}\( \frac{ \partial_\mu\phi \partial^\mu\phi}{f^2}\)^2 .
\end{align}

As presented above, these results  have been derived using a Pauli-Villars regularization scheme. Because this scheme requires massive fermions, this regulator explicitly breaks the chiral symmetry, and one may worry that these results we obtain in this way are an artifact of this breaking. However, we have verified that we find the same result  using  dimensional regularization. Because it is scale free, dimensional regularization does not reproduce the divergent terms and returns the finite pieces only. We further allay these concerns below in section \ref{sec:pathint}, where we demonstrate that this anomaly equation can be derived from the transformation of the path integral measure for a massless fermion.

%-------------------------------------------------------------
%-----Perturbative gravitational corrections------------------
%-------------------------------------------------------------

\section{Perturbative gravitational corrections}\label{sec:pertgrav}
\begin{figure}\label{fig:minkowskipertrules}
\centering
\includegraphics[scale = 0.8]{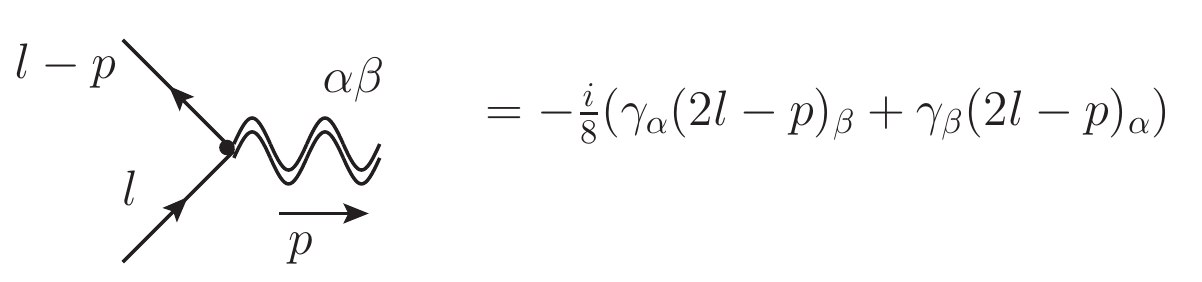}\\
\includegraphics[scale = 0.8]{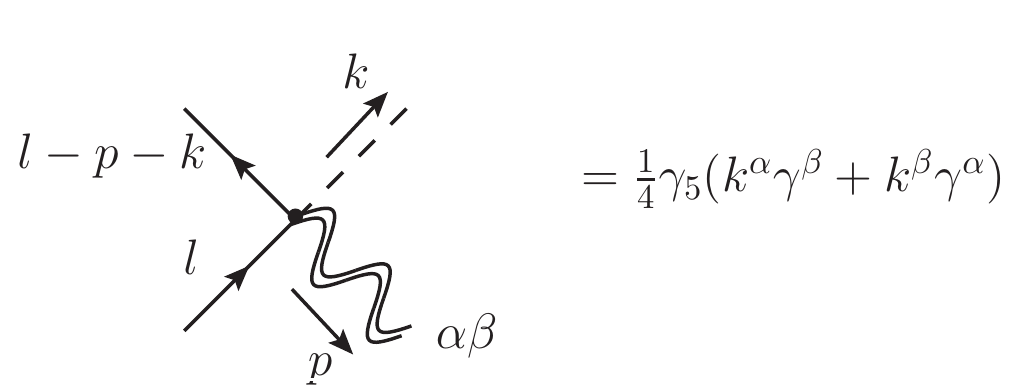}
\caption{Feynman rules for fermion-graviton, and axion-fermion-graviton interaction vertices}
\end{figure}

We next consider perturbative gravitational corrections to the amplitudes we computed above in section \ref{sec:minkowski}. We anticipate that additional contributions to the anomaly equation arise from the emission of gravitons from the processes above. We demonstrate that these corrections have two effects. First, we show that the gravitational corrections simply implement the minimal coupling procedure to take the partial derivatives to covariant derivatives as consistent with covariance. Second, we demonstrate that interactions with the gravitational field lead to new contributions to the anomaly equation. For simplicity we work only to linear order in the metric perturbation and with the single axion diagrams from above, inferring the non-linear results from covariance. We demonstrate in section \ref{sec:pathheat} below that our results are correct at the non-linear level.

%-------------------------------------------------------------
%-----Fermions in a linearized external gravitational field---
%-------------------------------------------------------------

\subsection{Fermions in a linearized external gravitational field}

In the presence of a gravitational field, the fermion action becomes
\begin{align}
S = \int {\rm d}^4x\; {\rm det}\, e \, \[  e^{\mu a}\frac{1}{2} \bar{\chi} i \gamma_a \dvec{D}_\mu \chi -m \bar{\chi}\chi + e^{\mu a} \frac{\partial_\mu\phi}{f} \bar{\chi}  \gamma_a \gamma_5\chi \],
\end{align}
where we have written the Lagrangian in an explicitly Hermitian form. Here $e^{\mu a}$ are the Vielbeins, defined
\begin{align}
g^{\mu\nu} = \eta_{ab}e^{\mu a}e^{\mu b},
\end{align}
where $g^{\mu\nu}$ is the spacetime metric, and $\eta_{ab}$ is the Minkowski metric. We study a metric $g_{\mu\nu}  =  \eta_{\mu\nu} + h_{\mu\nu}$, and work with the vierbein ${e}_{\mu a} = \eta_{\mu a} + \frac{1}{2}h_{\mu a}$. We treat the gravitational field as a general external source and for simplicity, we take the perturbation to be traceless.

There are three relevant gravitational interactions \cite{Delbourgo:1972xb, Delbourgo:1972en, Alvarez-Gaume:1983ihn}
\begin{align}\nn
\mathcal{L}_1 = & -\frac{1}{4}ih^{\mu\nu}\bar{\chi}\gamma_\mu \dvec{\partial}_\nu \chi , \quad 
\mathcal{L}_2 =   \frac{1}{16}i h_{\lambda\alpha}\partial_\mu h_{\nu\alpha}\bar{\chi}{\rm \Gamma}^{\mu\lambda\nu} \chi, \\
& \mathcal{L}_3 =  -\frac{1}{2}h^{\mu\nu}\frac{\partial_\mu \phi}{f} \bar{\chi}\gamma_\nu\gamma_5 \chi.
\end{align}
Note that the interaction $\mathcal{L}_2$ does not contain a term linear in the metric perturbation due to the anti-symmetry of ${\rm \Gamma}^{\mu\lambda\nu} = \{\gamma^\mu,[\gamma^\lambda, \gamma^\nu]\}$. Since we are only interested in one graviton contributions, we ignore this term in what follows. The Feynman rules for the interactions in $\mathcal{L}_1$ and $\mathcal{L}_3$ are shown in figure \ref{fig:minkowskipertrules}.

There are then two types of gravitational corrections to the anomaly equation. The first arise from the replacement of $\partial_\mu j_{5}^{\mu} \to g^{\mu\nu}\nabla_\mu j^e_{5 \nu}$, where $j^{e}_{5\nu}  = \bar{\chi} \gamma_5 e_{a\nu}\gamma^a \chi$, $j_{5\nu} = \eta_{a\nu}\bar{\chi} \gamma_5 \gamma^a \chi$, and the $\gamma^a$ are the Minkowski Dirac matrices. Expanding to linear order in the metric perturbation, we write this
\begin{align}\label{eqn:extgrav}
g^{\mu\nu}\nabla_\mu j^{e}_{5\nu}   = & i q_\mu j_{5\nu} \eta^{\mu\nu}-\frac{1}{2}\epsilon^{\mu\nu}(p) i (k_\mu + p_\mu) j_{5\nu},
\end{align}
where $\epsilon^{\mu\nu}(p)$ is the  polarization tensor for the gravitational field.  Using Pauli-Villars regularization, the one-axion part of $j_{5\nu}$ reads
\begin{align}
\langle j_{5\nu} \rangle_{\rm reg} = & \frac{i}{2\pi^2}  k_\nu \Bigg[ \sum c_i M_i^2 \(\log(M_i^2)\) +\frac{k^2}{6}\Bigg].
\end{align}

\begin{figure}
\centering
\includegraphics[scale = 0.8]{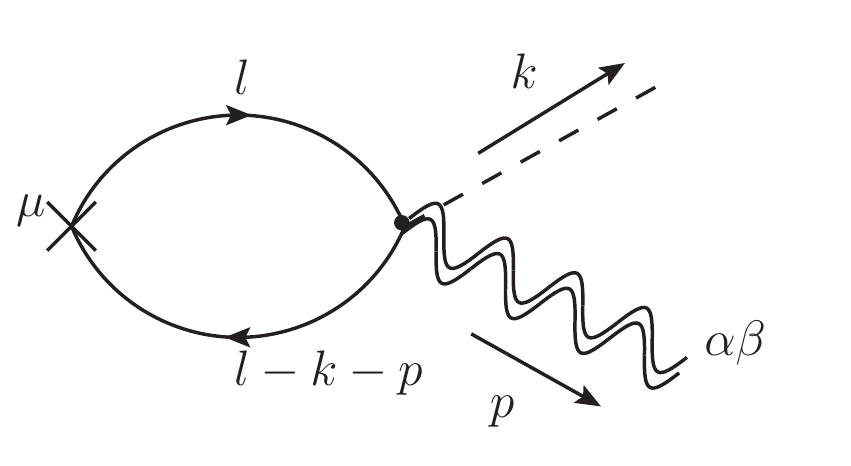}
\caption{One-vertex gravitational corrections to the one-axion contributions to the anomaly.}\label{fig:minkowskipertdiags_single}
\end{figure}

The second gravitational correction comes from graviton emission due to insertions of the vertices in figure \ref{fig:minkowskipertrules} when computing loop corrections to $j_{5\nu}$. The  gravitational corrections to the one-axion term in the left panel of figure \ref{fig:minkowskidiags} are shown in figures \ref{fig:minkowskipertdiags_single} and \ref{fig:minkowskipertdiags_tri}. We consider these separately in what follows.

%-------------------------------------------------------------
%-----1 vertex terms------------------------------------------
%-------------------------------------------------------------

\subsection{One-vertex terms}\label{sec:onevertexgrav}

We start with the graph that just has the one insertion of the axion-graviton vertex
\begin{align}
\int d^4 x e^{- i q\cdot x}\langle k, p | j^{\mu 5}(x)|0\rangle  = (2\pi)^4\delta^{(4)}(k+p-q) \epsilon^{\alpha\beta}(p) \mathcal{M}_{1,\alpha\beta}^{\mu}(k, p),
\end{align}
where we regulate as above with Pauli-Villars fields
\begin{align}\nn
\mathcal{M}_{1, \alpha\beta, \rm reg}^\mu(k, p) = & - \int \frac{d^4 l}{(2\pi)^4}{\rm tr}\Bigg[\gamma^\mu \gamma_5 \frac{i((\slashed{l}-\slashed{k}-\slashed{p})+m)}{(l - k-p)^2-m^2}\(\frac{1}{4} \gamma_5 (k_\alpha \gamma_\beta+k_\beta \gamma_\alpha )\) \frac{i(\slashed{l}+m)}{l^2-m^2}\\
& +\sum c_i(m \to M_i)\Bigg].
\end{align}
We compute the divergence of the axial current, and use
\begin{align}
q_\mu \gamma^\mu \gamma_5 = (\slashed{l}-m)\gamma_5 + \gamma_5 (\slashed{l}-\slashed{k}-\slashed{p}-m) + 2\gamma_5 m,
\end{align}
and now because the integrals are finite, we can safely shift the momenta to find
\begin{align}\nn
iq_\mu \mathcal{M}_{1, \alpha\beta, \rm reg}^\mu(k) = & \frac{i}{2}\int \frac{d^4 l}{(2\pi)^4}{\rm tr}\Bigg[ m \gamma_5 \frac{(\slashed{l}-\slashed{k}-\slashed{p}+m)}{(l - k-p)^2-m^2}  \gamma_5 \(k_\alpha \gamma_\beta+k_\beta \gamma_\alpha\) \frac{\slashed{l}+m}{(l^2-m^2)}\\ & +\sum_i c_i (m\to M_i)\Bigg].
\end{align}
The integral can be performed using the standard methods, and the result in the limit $m \to 0$, $M_i \to \infty$ is
\begin{align}
iq_\mu \mathcal{M}^\mu_{1, \alpha\beta, \rm reg}(k)= & \frac{1}{2\pi^2} \left(({k}+{p})_\alpha k_\beta+({k}+{p})_\beta k_\alpha\right) \[ \sum_i c_i M_i^2 \log\[M_i^2\] +\frac{1}{6}(k+p)^2 \].
\end{align}
Again, we find a divergent contribution. As above, we show that this can be absorbed by renormalizing the axion kinetic term. Next we compute the triangle graph terms.

%-------------------------------------------------------------
%-----2 vertex terms------------------------------------------
%-------------------------------------------------------------

\subsection{Two-vertex terms}\label{sec:twovertexgrav}

\begin{figure}
\centering
\includegraphics[scale = 0.8]{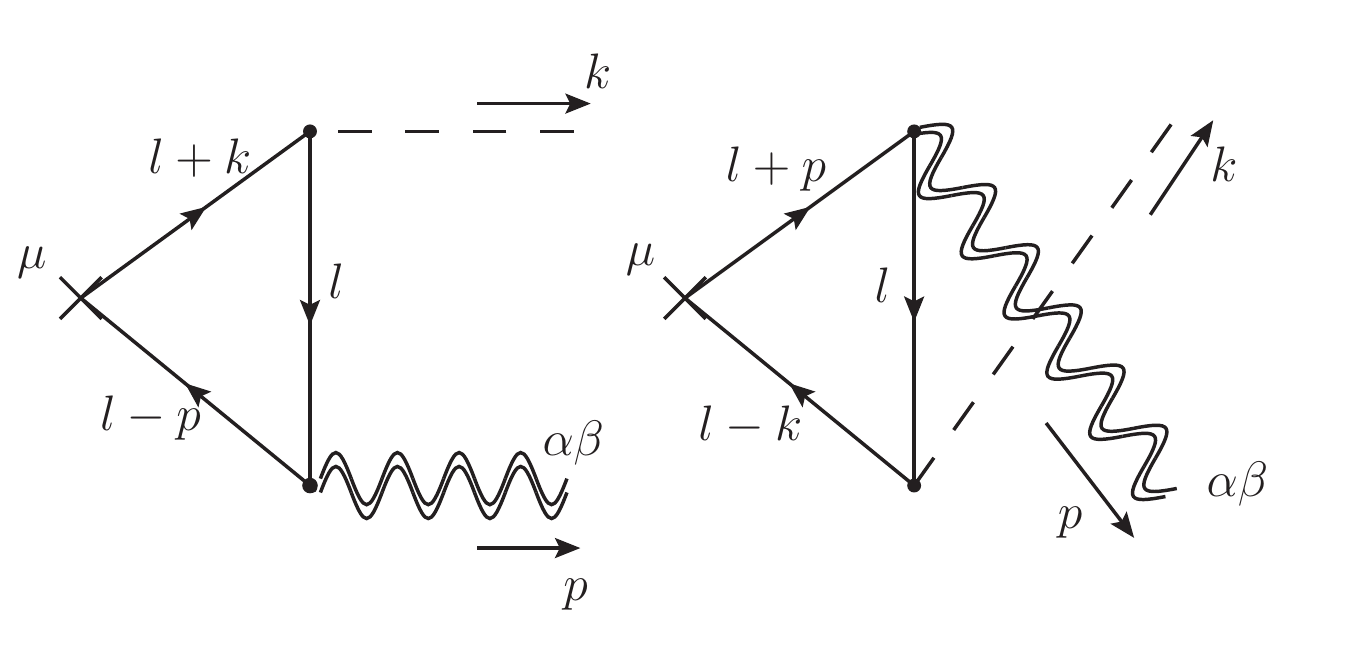}
\caption{Two-vertex gravitational corrections to the one-axion contribution to the anomaly.}\label{fig:minkowskipertdiags_tri}
\end{figure}
We now consider the terms that arise from diagrams with an axion-fermion vertex, and a fermion-graviton vertex as displayed in figure \ref{fig:minkowskipertdiags_tri}. This diagram leads to a contribution to the divergence of the axial current with a final state graviton and axion
\begin{align}
\int d^4 x e^{- i q\cdot x}\langle k ,p| j^{\mu 5}(x)|0\rangle  = (2\pi)^4\delta^{(4)}(k+p-q) \epsilon^{\alpha\beta}(p)\mathcal{M}^\mu{}_{2, \alpha\beta}(k, p),
\end{align}
where as above we regularize using Pauli-Villars fields, and 
\Beq
\mathcal{M}_{2,\alpha\beta, \rm reg}^\mu&(k, p)=\\
&- \frac{i}{8} \int \frac{d^4 l}{(2\pi)^4}  {\rm tr}\Bigg[\gamma^\mu \gamma_5 \frac{i((\slashed{l}-\slashed{p})+m)}{(l - p)^2-m^2}\gamma_\alpha(2l-p)_\beta\frac{i(\slashed{l}+m)}{l^2-m^2}\gamma_5\slashed{k}\frac{i((\slashed{l}+\slashed{k})+m)}{(l +k)^2-m^2} \\
& 
+ \gamma^\mu \gamma_5 \frac{i((\slashed{l}-\slashed{k})+m)}{(l - k)^2-m^2}\gamma_5 \slashed{k} \frac{i(\slashed{l}+m)}{l^2-m^2}\gamma_\alpha(2l+p)_\beta\frac{i((\slashed{l}+\slashed{p})+m)}{(l + p)^2-m^2} +\alpha \leftrightarrow \beta\\
& +\sum c_i(m \to M_i)\Bigg].
\Eeq
Taking the divergence of the axial current is equivalent to dotting this with $iq_{\mu}$,  we write
\Beq
q_\mu \gamma^\mu \gamma_5 = & (\slashed{l}+\slashed{k}-m)\gamma_5 + \gamma_5 (\slashed{l}-\slashed{p}-m) + 2\gamma_5 m,\\ 
q_\mu \gamma^\mu \gamma_5 = & (\slashed{l}+\slashed{p}-m)\gamma_5 + \gamma_5 (\slashed{l}-\slashed{k}-m) + 2\gamma_5 m,
\Eeq
which, after shifting momenta puts this in the simpler form
\begin{align}\nn
&iq_\mu  \mathcal{M}_{2, \alpha\beta, \rm reg}^\mu(k, p)=\\\nn
&- \frac{i}{4} \int \frac{d^4 l}{(2\pi)^4}  {\rm tr}\Bigg[\gamma^\mu \gamma_5 \(\gamma_\alpha(k+p)_\beta+\gamma_\beta(k+p)_\alpha \) \frac{(\slashed{l}+m)}{l^2-m^2}\gamma_5\slashed{k}\frac{((\slashed{l}+\slashed{k})+m)}{(l +k)^2-m^2} \\\nn
& + m \gamma_5 \frac{((\slashed{l}-\slashed{p})+m)}{(l - p)^2-m^2}\(\gamma_\alpha(2l-p)_\beta+\gamma_\beta(2l-p)_\alpha\) \frac{(\slashed{l}+m)}{l^2-m^2}\gamma_5\slashed{k}\frac{((\slashed{l}+\slashed{k})+m)}{(l +k)^2-m^2} \\ \nn
& 
+ m\gamma_5 \frac{((\slashed{l}-\slashed{k})+m)}{(l - k)^2-m^2}\gamma_5 \slashed{k} \frac{(\slashed{l}+m)}{l^2-m^2}\(\gamma_\alpha(2l+p)_\beta+\gamma_\beta(2l+p)_\alpha\)\frac{((\slashed{l}+\slashed{p})+m)}{(l + p)^2-m^2} \\
& +\sum c_i(m \to M_i)\Bigg].
\end{align}
This expression can be evaluated using the standard techniques to obtain, in the limit $m\to 0$, $M_i \to\infty$  \Beq
 i q_\mu &\mathcal{M}_{2, \alpha\beta, \rm reg}^\mu(k, p) 
        =\\
        &\frac{1}{(4\pi)^2} \frac{1}{6}\Bigg[ 4 k_{\alpha } k_{\beta }
   \left(2k^{2 } +p^2 +2 k^{\delta } p_{\delta }\right)+2(k_{\alpha }p_{\beta } +p_{\alpha} k_{\beta } )\big(2 k^2+p^2\big)+4
   k^2 p_{\alpha } p_{\beta }\Bigg].
\Eeq

%-------------------------------------------------------------
%-----Total perturbed Minkowski result------------------------
%-------------------------------------------------------------

\subsection{Total perturbed Minkowski result}

Putting together the results from  eq.\ \ref{eqn:extgrav} and sections \ref{sec:onevertexgrav} and \ref{sec:twovertexgrav}, we find the total result for the one-axion one-graviton contribution to the divergence of the axial current can be written
\begin{align}\nn \label{eqn:pertanom}
\langle\nabla_\mu j_{5}^\mu\rangle_{\rm reg}%
 & = -\frac{1}{2 \pi^2} \(k^2  -(k_\alpha + p_\alpha)k_\beta \epsilon^{\alpha\beta}(p) \) \sum c_i M_i^2 \log(M_i^2)\\\nn
& - \frac{1}{12\pi^2}\(k^4  -((k+p)^2+k^2) (k_\alpha +p_\alpha)k_\beta   \epsilon^{\alpha\beta}(p) \)\\
& + \frac{1}{24\pi^2}\Big(p^2k_\beta  k_\alpha   -2 k^\nu p_\nu p_\alpha k_\beta+k^2 p_\alpha p_\beta \Big)\epsilon^{\alpha\beta}(p).
\end{align}
Then using the linearized Riemann tensor
\begin{align}
R_{\alpha\beta\mu\nu} & = \frac{1}{2}\[\partial_{\beta}\partial_{\mu}h_{\alpha\nu} + \partial_{\alpha}\partial_{\nu}h_{\beta\mu} -\partial_{\alpha}\partial_{\mu}h_{\beta\nu}- \partial_{\beta}\partial_{\nu}h_{\alpha\mu}\],
\end{align}
eq.\ \eqref{eqn:pertanom} implies, 
\Beq
\nabla_\mu j^{\mu}_5&=\frac{\Box \phi}{2\pi^2f}\sum_i c_i M_i^2 \log\[M_i^2\] -\frac{\Box^2 \phi}{12\pi^2f}-\frac{\nabla_\mu\(G^{\mu\nu}\partial_\nu\phi\)}{12\pi^2f}+\frac{1}{3\pi^2}\nabla_\mu \left(\frac{\partial^\mu\phi}{f}\frac{\partial_\nu\phi}{f}\frac{\partial^\nu\phi}{f}\right).
\Eeq
Here $G^{\mu\nu}$ is the Einstein tensor, $\Box = \nabla^\alpha \nabla_\alpha$, and $\nabla_\alpha$ is the covariant derivative. Note that we have only shown this for the one-axion terms, and only to linear order in perturbations around Minkowski spacetime. However, general covariance dictates that the full non-linear anomaly equation should be of this form. We prove this using the path integral below in section \ref{sec:pathheat}. In the next section, we show that this gravitational result, obtained perturbatively about Minkowski space holds about a de Sitter background.

%-------------------------------------------------------------
%-----Axion anomalies in de Sitter space----------------------
%-------------------------------------------------------------

\section{Axion anomalies in de Sitter space}\label{sec:desitter}

In this section we verify that the Minkowski space anomaly equation we have found above is satisfied about a de Sitter background through direct computation, including the gravitational terms. We consider a fermion propagating on a rolling, classical axion background in de Sitter space, and compute the divergence of the axial vector current on this background. We regularize our result in two ways, first using Pauli-Villars fields analogously to our Minkowski results, and second we implement adiabatic subtraction.

%-------------------------------------------------------------
%-----Fermions during pseudoscalar-driven inflation-----------
%-------------------------------------------------------------

\subsection{Fermions during pseudoscalar-driven inflation}

We consider the theory of a pseudoscalar inflaton, $\phi$, coupled to a single generation of Dirac  fermions, $\chi$, described by the action \cite{Adshead:2015kza, Adshead:2015jza, Adshead:2018oaa, Adshead:2019aac}
\begin{align}
S = \int {\rm d}^4 x \sqrt{-g}\[\frac{M_{\rm Pl}^2}{2}R + \frac{1}{2}(\partial\phi)^2 - V(\phi) + i \bar\chi e^{\mu}{}_{a}\gamma^{a}\nabla_{\mu} \chi-m\bar{\chi}\chi+\frac{\partial_{\mu}\phi}{f} \bar{\chi}e^{\mu}{}_{a}\gamma^{a}\gamma_5\chi\],
\end{align}
where $R$ is the Ricci scalar, $\mu, \nu, \ldots$ are space-time indices, and $a, b,\ldots$ are Lorentz indices. We work with the conformal Friedmann-Robertson-Walker metric
\Beq
{\rm d} s^2=a^2({\rm d}\tau^2-\delta_{ij} {\rm d} x^i {\rm d} x^j)
\Eeq
where we ignore metric fluctuations. We work in the de Sitter limit, and approximate the motion of the axion as rolling at a constant rate in cosmic time $t$. With these approximations
\begin{align}
a = e^{H t} = \frac{1}{-H \tau}, \quad H = \frac{1}{a}\frac{d a}{dt}, \quad \frac{d{\phi}}{dt} = \frac{1}{a}\frac{d\phi}{d\tau} = {\rm const}.
\end{align}
On the de Sitter background the Vierbiens and spin connection are
\begin{align}
e_{\mu}{}^{a} = a \delta_{\mu}{}^{a}, \quad e^{\mu}{}_{a} = a^{-1}\delta^{\mu}{}_a, \quad e_{\mu a} = a \eta_{\mu a}, \quad \omega_{\mu a b }= &{\cal H}( \delta^{0}{}_{b} \eta_{\mu a} - \delta^0{}_{a} \eta_{\mu b} ),
\end{align}
where $ \eta_{\mu\nu}$ is the Minkowski metric and $\mathcal{H} = \partial_\tau a/a$. It proves useful to rescale the fields $\psi = a^{3/2} \chi$, to obtain
\begin{align}
S = \int {\rm d}^4 x \[i\bar{\psi} \gamma^{\mu}\partial_{\mu} \psi-a m \bar{\psi}\psi+\frac{\partial_{\mu}\phi }{f}\bar{\psi}\gamma^\mu\gamma_5 \psi \].
\end{align}
We expand the fields into modes,
\begin{align}
\psi({\bf k}, \tau) = & \sum_{r = \pm} (U_r({\bf k}, \tau) b^r_{\bf k}+V_r(-{\bf k}, \tau) c^{r,\dagger}_{-\bf k}),%\\
\end{align}
where the creation and annihilation operators satisfy the anti-commutation relations
\begin{align}
\{b^{\lambda}_{\bf k}, b^{\lambda' \dagger}_{\bf p}\}= &(2\pi)^3\delta_{\lambda\lambda'}\delta^{3}({\bf k}-{\bf p}), \quad 
\{c^{\lambda}_{\bf k}, c^{\lambda' \dagger}_{\bf p}\} =  (2\pi)^3\delta_{\lambda\lambda'}\delta^{3}({\bf k}-{\bf p}),
\end{align}
with all other anti-commutators vanishing. Charge conjugation relates $V$  to $U$, where
\begin{align}
U_r({\bf k}, \tau) = \frac{1}{\sqrt{2}}\(\begin{matrix} u_r(k, \tau)\\ r v_{r}(k, \tau) \end{matrix}\)\chi_r({\bf k}), \quad V_r({\bf k}, \tau) = &\frac{1}{\sqrt{2}}\(\begin{matrix} v^*_{r}(k, \tau)\\ -r u^*_{r}(k, \tau) \end{matrix}\)\chi_{-r}({\bf k}),
\end{align}
and the helicity-$r$ two spinors, $\chi_r({\bf k})$, satisfy $\sigma\cdot{\bf k}\chi_r({\bf k})=rk\chi_r({\bf k})$ and are normalized via
\begin{align}
\chi^\dagger_r({\bf k})\chi_s({\bf k}) = \delta_{rs}.
\end{align}
Explicitly, we use the Dirac matrices,
\begin{align}
\gamma^0 = &\( \begin{matrix}  \mathds{1} & 0 \\0 & - \mathds{1} \end{matrix}\), \quad \gamma^i = \( \begin{matrix} 0 &  \sigma^i \\  - \sigma^i & 0 \end{matrix}\), \quad  \gamma_5 = \( \begin{matrix} 0 & \mathds{1} \\ \mathds{1}&  0 \end{matrix}\),
\end{align}
and define 
\begin{align}
\mu\equiv \frac{m}{H}\,,\quad\xi\equiv -\frac{\dot\phi_0}{2\,f\,H}\,,\quad x\equiv -k\tau \,. 
\end{align}
The Dirac equation then gives the following system 
\begin{eqnarray} 
\frac{\partial  u_r}{ \partial x}&=& i \, \frac{\mu}{x} \,  u_r+i\,\left( 1 +  \frac{2\xi}{x}  \,r\right)  \, v_r\,,\quad  \frac{\partial  v_r}{ \partial x}=- i \, \frac{\mu}{x}   \,  v_r+i\,\left( 1+\frac{2 \xi}{x}\,r\right)  \, u_r\,.
\end{eqnarray}
The system is solved by
\begin{equation}
 u_r = \frac{1}{ \sqrt{2 x}} \left( s_r + d_r \right) \;\;,\;\; 
 v_r = \frac{1}{ \sqrt{2 x}} \left( s_r - d_r \right) \,,  
\label{eq:def_uvtilde} 
\end{equation}   
with
\begin{eqnarray}\label{eqn:dssols}
 s_r =   {\rm e}^{-\pi r \xi} \, {W}_{\frac{1}{2}+2 i r \xi ,\, i \sqrt{\mu^2+4 \xi^2}}(- 2 i x)\,,\quad 
 d_r= - i  \, \mu \, {\rm e}^{-\pi r \xi} \,W_{-\frac{1}{2}+2 i r \xi ,\, i \sqrt{\mu^2+4 \xi^2}}(- 2 i x)\,, 
\end{eqnarray}
where ${W}_{\alpha,\beta}(z)$ denotes the  Whittaker W-function. The integration constants have been determined by imposing the normalization $\vert  u_r \vert^2 + \vert  v_r \vert^2 = 2$ and the positive frequency condition
\begin{align}
\lim_{x \to \infty} u_r \left( x \right) =  \lim_{x \to \infty} v_r \left( x \right) = 
{\rm e}^{i \left( x + 2 r \xi \ln \left( 2 x \right) - \frac{\pi}{4} \right) } .
\label{norma-uv} 
\end{align}
We can make use of these solutions to compute the divergence of the axial current in the axion background in de Sitter space. We are interested in the expectation value of 
\begin{align}
\nabla_\mu j^{\mu}_5 = 2 m i \bar{\psi}\gamma_5 \psi,
\end{align}
where we used the classical equation of motion. We evaluate the right-hand side and then take the limit $m\to 0$. Taking the expectation value with the Bunch-Davies vacuum, we get
\begin{align}\label{eqn:axialcurrentdS}
\langle \nabla_\mu j^\mu_5 \rangle  &= \dfrac{m}{ a^3(\tau)} \int \dfrac{{\rm d}^3p}{(2\pi)^3} \sum_r
[-ir ( u^*  v -  v^*  u)]_{r,p,\tau} =2 m H^3  \dfrac{1}{2 \pi^2} \sum_r r \int {\rm d} y\,y \,\Im(d^*_r s_r) \,, 
\end{align}
where $d_r$ and $s_r$ are the functions given in eq.\ \eqref{eqn:dssols}. This integral is divergent, and must be regularized. In what follows, we discuss two different regularization schemes, Pauli-Villars regularization as used above in Minkowski space, and adiabatic regularization.

%-------------------------------------------------------------
%-----Pauli-Villars regularization----------------------------
%-------------------------------------------------------------

\subsection{Pauli-Villars regularization}

In order to regularize the divergent integral for the current, we introduce a set of Pauli-Villars fields so that the expectation value in eq.\ \eqref{eqn:axialcurrentdS} becomes
\begin{align}\label{eqn:axialcurrentdS1}
\langle \nabla_\mu ( j^\mu_5 +\sum_i  c_i j^{\mu}_{5,i})\rangle  = &  \frac{2 i}{a^3}\(   m  \langle \bar{\psi} \gamma_5 \psi\rangle + \sum_i c_i  M_{i} \langle \bar{\psi}_i  \gamma_5 \psi_i\rangle\).
\end{align}

Somewhat remarkably,  the integral on the right-hand-side of eq.\ \eqref{eqn:axialcurrentdS1} can be done analytically with  the result  \cite{Adshead:2018oaa}
\begin{align}\nn \label{eq:axialexact}
\langle \nabla_\mu &( j^\mu_5 +\sum_i c_i j^{\mu}_{5,i})\rangle= \\
  &  2 \mu ^2H^4\dfrac{1}{2 \pi^2}
\left[ -6 \xi \gamma_E-\frac{3}{2} \sqrt{\mu ^2+4 \xi ^2} \sinh (4 \pi  \xi ) \text{csch}\left(2 \pi  \sqrt{\mu ^2+4 \xi ^2}\right)+7 \xi \right. \nonumber \\
&  \left. + \frac{1}{2}\Re\Bigg\{ i \left(\mu ^2- 8 \xi^2 - 6 i \xi +1\right) 
\left[H_{-i \left(2 \xi +\sqrt{\mu ^2+4 \xi ^2}\right)} \left(\sinh (4 \pi  \xi ) \text{csch}\left(2 \pi  \sqrt{\mu ^2+4 \xi ^2}\right)+1\right)
\right. \right. \nonumber \\
& \qquad \qquad \left. \left.
+H_{i \left(\sqrt{\mu ^2+4 \xi ^2}-2 \xi \right)} \left(1-\sinh (4 \pi  \xi ) \text{csch}\left(2 \pi  \sqrt{\mu ^2+4 \xi ^2}\right)\right)\right] \Bigg\} \right] \nonumber \\
& \qquad \qquad +\frac{1}{2\pi^2}\[\xi H^4 \(16 \mu ^2-16 \xi ^2+3\right)- 6 \xi  H^4  \sum_ic_i  \mu_i ^2 \log \left(\mu_i \right)\],
\end{align}
where $H_n$ is the harmonic number. In the  limit $\mu \ll 1$, eq.\ \eqref{eq:axialexact} is 
\begin{align}
\langle \nabla_\mu ( j^\mu_5 +\sum_i c_i j^{\mu}_{5,i})\rangle 
  = &  \dfrac{H^4}{2 \pi^2}\[-8 \pi\mu^2 \, \xi^2 +\xi  \(16 \mu ^2-16 \xi ^2+3\right)-6 \xi   \sum_i c_i  \mu_i^2 \log \left(\mu_i \right)\],
\label{eq:PV_axial}
\end{align}
and the result does not vanish when the physical fermion mass vanishes, $\mu = m/H \rightarrow 0$. Note that, in the limit where the pseudoscalar rolls at a constant rate in cosmic time, and the spacetime is de Sitter we can evaluate 
\Beq
\Gamma^0{}_{ij}=Ha^2\delta_{ij}\,,\quad G^{00}=3H^2\, ,\quad  \Box\phi = 3H\dot{\phi}, \quad \Box^2\phi = 0.
\Eeq
Inserting these into the result from  eq.\ \eqref{eqn:heatanomy} above, the anomaly equation is
\Beq
\langle \nabla_\mu j^{\mu}_{5}(x)\rangle&=-\frac{1}{4\pi^2}\frac{\dot{\phi}}{Hf}\left[3-4\left(\frac{\dot{\phi}}{fH}\right)^2\right]H^4=\frac{1}{2\pi^2}\xi \left[3-16 \xi^2\right]H^4,
\Eeq
which agrees with our computation. We have also verified that this result can be obtained perturbatively in de Sitter space using the in-in formalism and computing diagrams analogous to those in section \ref{sec:minkowski}. 
%-------------------------------------------------------------
%-----Adiabatic regularization--------------------------------
%-------------------------------------------------------------

\subsection{Adiabatic regularization}

We can check that our result is independent of regularization method. Here we show the result for adiabatic subtraction. This method subtracts the adiabatic solutions mode by mode from the integrand. Because it is a scale free regularization scheme, power-law divergences are discarded by this method. 

We begin with the equations \cite{Adshead:2018oaa} 
\begin{align}
i \dot{ u}_r &= m  u_r + \left( \dfrac{k}{a} + 2 H r \xi \right)  v_r, \quad 
i \dot{ v}_r = - m   v_r + \left( \dfrac{k}{a} + 2 H r \xi \right)   u_r,
\end{align}
which we solve as an expansion in the Hubble parameter $H$.  We introduce the ansatz  \cite{Landete:2013axa, Landete:2013lpa, delRio:2014cha}
\begin{align}
 u_r &= \sqrt{ 1 + \dfrac{m}{\omega}} \exp \left(
-i \int^t d\tilde{t} \left[ \sum_{n = 0}^{\infty} H^{n} \omega_n (\tilde{t}) \right] \right)
\left[ 1 + \sum_{m = 1}^{\infty }H^m F_{m,r} \right], \nonumber \\
 v_r &= \sqrt{ 1 - \dfrac{m}{\omega}} \exp \left(
-i \int^t d\tilde{t} \left[\sum_{n = 0}^{\infty} H^{n} \omega_n (\tilde{t}) \right] \right)
\left[ 1 + \sum_{m = 1}^{\infty }H^m G_{m,r} \right].
\end{align}
We can then compute the adiabatic part of the axial current, to fourth order in powers of $H$,\footnote{While naively the total axial current is made finite by subtracting only the terms to quadratic order in $H$, the situation here is analogous to the computation of the trace anomaly (see, for example, \cite{Parker:2009uva}). There the individual terms in the trace anomaly are more divergent that their sum, and one needs to keep up to fourth order to reproduce the anomaly. In this case, the contribution of each helicity to the axial current is much more divergent their sum due to cancellations.} we have
\begin{align}\label{eqn:adiabaticreg}
\langle \nabla_\mu J^\mu_5 \rangle_{\rm Ad.}    & =  \dfrac{m}{ a^3(\tau)} \int \dfrac{{\rm d}^3p}{(2\pi)^3} \sum_r
[-ir ( u^*  v -  v^*  u)]_{r,p,\tau} \\\nn
 &= \dfrac{2 m}{ a^3} \int \dfrac{{\rm d}^3p}{(2\pi)^3} \sum_r r
\dfrac{p/a}{\omega} \Bigg[ H^2( \Im\left[   G_{2,r}\] +  F_{1,r}\Im\left[  G_{1,r}\])\\ \nn & 
\qquad + H^4( \Im\[ G_{4,r}\right]+ F_{1,r}\Im\left[ G_{3,r}\right]
+  F_{2,r}\Im\left[G_{2,r} \right]  + F_{3,r} \Im\left[ G_{1,r}\right])\Bigg]_{r,p,\tau} \, +\mathcal{O}(H^5)
\end{align}
then substituting in the adiabatic solutions from Appendix \ref{app:adiabatic} and performing the integration and taking the limit $\mu = m/H\ll1$, we obtain
\begin{align}\label{eqn:fulladreg1}
\langle \nabla_\mu j^\mu_5 \rangle_{\rm reg.} = \langle \nabla_\mu j^\mu_5 \rangle-\langle \nabla_\mu j^\mu_5 \rangle_{\rm Ad.}  
&= \dfrac{H^4}{2 \pi^2}\[-8 \pi\mu^2 \, \xi^2 +\xi  \(16 \mu ^2-16 \xi ^2+3\right)\],
\end{align}
and again we match the anomaly equation.

%-------------------------------------------------------------
%-----Anomaly from the Path Integral Measure------------------
%-------------------------------------------------------------

\section{Transformation of the path integral measure}\label{sec:pathint}

In moving from the basis in eq.\ \eqref{eqn:fermaxaction} to the basis in eq.\ \eqref{eqn:fermaxaction2}, a chiral rotation of the fermion field is performed. It is well known that the path integral measure of fermion theories involving gauge and gravitational fields is not invariant under chiral rotations. In this section, we demonstrate that the measure is not invariant in the presence of an axion field.  

We study the transformation properties of the measure of the path integral under the change of basis in eq.\ \eqref{eqn:axialrot}  using Fujikawa's Euclidean path integral method \cite{Fujikawa:1979ay, Fujikawa:1980eg}. We perform this computation in two ways, the first by expanding the anomaly function into momentum states, and the second by recasting the anomaly function as the solution to the heat equation. We show that the chiral rotation that transforms the action from the form in eq.\ \eqref{eqn:fermaxaction} to eq.\ \eqref{eqn:fermaxaction2} generates Jacobian terms that exactly reproduce the anomaly equation.

%-------------------------------------------------------------
%-----Preliminaries-------------------------------------------
%-------------------------------------------------------------

\subsection{Preliminaries}

We consider the Euclidean path integral for a Dirac fermion interacting with an axion field
\begin{align}
\mathcal{Z} = \mathcal{N} \int {\rm d}\bar\psi {\rm d}\psi e^{-S}, \quad S = \int d^4 x\(\bar{\psi}\slashed{D}\psi\),
\end{align}
where in Euclidean space, the Dirac matrices satisfy
\begin{align}
\{\gamma_\mu, \gamma_{\nu}\} = 2\delta_{\mu\nu}, \quad \{\gamma_5, \gamma_\mu\} = 0,\quad \gamma_5^\dagger = \gamma_5, \quad {\rm Tr}\gamma_{5} \gamma_\mu\gamma_\nu\gamma_\sigma\gamma_\tau = 4\epsilon_{\mu\nu\sigma\tau}.
\end{align}
The covariant derivative encodes the interactions with the gauge, gravitational, and axion fields\footnote{Here for notational simplicity, we have included the axion into the covariant derivative.} 
\begin{align}\label{eqn:diraccov}
\slashed{D}&=i\gamma^ae^\mu_a\left(\partial_\mu+\frac{1}{8}[\gamma_b,\gamma_c]\sigma_\mu^{bc}+A_\mu+i\gamma_5\frac{\partial_\mu\phi}{f}\right),
\end{align}
and the spin connection is
\begin{align}
\sigma^{ab}_\mu&=e^{a\nu}\Gamma^\rho_{\mu\nu}e^b_\rho-e^{a\nu}\partial_{\mu}e^b_\nu.
\end{align}
After the appropriate rotations to Euclidean space, the operator in eq.\ \eqref{eqn:diraccov} is self-adjoint, and its eigenspinors form an orthonormal basis
\Beq\label{eqn:diracop}
\slashed{D}\psi_n(x)=\lambda_n\psi_n(x),\qquad \int d^4x\sqrt{g}\bar{\psi}_n(x){\psi}_m(x)=\delta_{nm}\,,
\Eeq
where $\lambda_n$ is real. We expand the Dirac field into the eigenspinors
\Beq
\label{eq:DiracExpansion}
\psi(x)=\sum_n a_n\psi_n(x)\,,\quad 
\bar{\psi}(x)=\sum_n \bar{b}_n\bar{\psi}_n(x)\,,
\Eeq
where $a_n$ and $b_n$ are Grassmann-valued numbers. Under local chiral or axial rotations
$
\psi(x) \rightarrow e^{i\gamma_5\alpha(x)}\psi(x)\,,
$
the action transforms as
\Beq
S&\rightarrow S-\int d^4x \sqrt{g}\, \nabla_\mu j_5^{\mu}\alpha(x) \,.
\Eeq
For an infinitesimal local chiral rotation, $|\alpha(x)|\ll1$,
\Beq
\delta\psi(x)=i\alpha(x)\gamma_5\psi(x)\,,\quad \delta\bar{\psi}(x)=i\alpha(x)\bar{\psi}(x)\gamma_5\,,
\Eeq
which implies
\begin{align}
\delta a_n(x)&=\sum_m X_{nm} a_m(x)\,,\quad \delta b_n(x)=\sum_m X_{nm} \bar{b}_m(x)\,,
\end{align}
where
\begin{align}
X_{nm}&=i\int d^4x\sqrt{g}\alpha(x)\bar{\psi}_n(x)\gamma_5{\psi}_m(x)\,.
\end{align}
Therefore, under an infinitesimal chiral rotation the path integral measure transforms as
\Beq
\mathcal{D}\bar{\psi}\mathcal{D}\psi=\Pi_n {\rm d} a_n  {\rm d}  \bar{b}_n\rightarrow \Pi_n  {\rm d}  a_n  {\rm d}  \bar{b}_n {\rm det}^{-2}(1+X)\,.
\Eeq
We write
\Beq
{\rm det}^{-2}(1+X)\approx {\rm det}(1-2X)\approx {\rm det}(e^{-2X})\approx e^{-2{\rm Tr}X}\,.
\Eeq
so that the effect of the local chiral rotation is to shift the partition function
\Beq
\label{eq:Z}
\mathcal{Z}\rightarrow \left[1+\int d^4x\sqrt{g}\left[\left(\langle \nabla_\mu j_5^{\mu}(x)\rangle-2i\mathscr{A}(x)\right)\alpha(x)+\mathcal{O}(\alpha^2(x))\right]\right]\mathcal{Z},
\Eeq
where 
\begin{align}\label{eqn:anomfunct}
\mathscr{A}(x) = \sum_n\bar{\psi}_n(x)\gamma_5{\psi}_n(x),
\end{align}
is the anomaly function. The local chiral rotation can be interpreted as an infinitesimal change of the $\psi$ variables in the path Integral, implying that $\mathcal{Z} \rightarrow \mathcal{Z}$. Therefore (since the infinitesimal parameter $\alpha(x)$ is arbitrary)
\Beq
\label{eq:AxAnom}
\langle \nabla_\mu j_5^{\mu}(x)\rangle=2i\mathscr{A}(x).
\Eeq
The sum in eq.\ \eqref{eqn:anomfunct} is ill defined and must be regulated. We regulate it using the eigenvalues of the Dirac operator in eq.\ \eqref{eqn:diracop}, inserting a factor of  $\exp(-\lambda^2_n/M^2)$ into the sum
\Beq
\label{eq:reg}
\mathscr{A}(x) =\lim_{M \rightarrow \infty}\sum_n\bar{\psi}_n(x)\gamma_5e^{-\frac{\lambda^2_n}{M^2}}{\psi}_n(x) =\lim_{M \rightarrow \infty}\sum_n\bar{\psi}_n(x)\gamma_5e^{-\frac{\slashed{D}^2}{M^2}}{\psi}_n(x).
\Eeq
We now evaluate the sum in two different ways.

%-------------------------------------------------------------
%-----Momentum space------------------------------------------
%-------------------------------------------------------------

\subsection{Momentum space}\label{sec:pathmom}

We begin by simply evaluating the trace by following the approach that can be found in textbooks---by expanding into a basis of momentum states (see, for example, Ref.\ \cite{Peskin:1995ev}). For simplicity, we ignore the gravitational field and the gauge field. We restore these in section \ref{sec:pathheat} below.

We seek to evaluate
\begin{align}
\lim_{M \rightarrow \infty}\sum_n\bar{\psi}_n(x)\gamma_5e^{-\frac{\slashed{D}^2}{M^2}}{\psi}_n(x)= \lim_{M\to \infty} {\rm Tr}\langle x |\gamma_5 e^{\slashed{D}^2/M^2}|x\rangle
\end{align}
where the trace runs over the Dirac indices, and
\begin{align}
\slashed{D}_x^2 & = \Box + \frac{(\partial\phi)^2}{f^2} +i\frac{\gamma_5}{f} \Box\phi- i \gamma_5 \[\gamma^\mu, \gamma^{\nu}\] \frac{ \partial_\mu\phi}{f} \partial_\nu .
\end{align}
We evaluate the trace by inserting a complete set of momentum states
\begin{align}\nn
&\lim_{M\to \infty}  {\rm tr}\langle x |\gamma_5 e^{\slashed{D}^2/M^2}|x\rangle  = \lim_{M\to \infty} {\rm tr} \int \frac{{\rm d}^4 k}{(2\pi)^4}\langle x | e^{\slashed{D}^2/M^2}|k\rangle\langle k |x\rangle\\\nn
& =  i\lim_{M\to \infty} \int \frac{{\rm d}^4 k}{(2\pi)^4} e^{-\frac{k^2}{M^2}}{\rm tr}\Bigg[ \frac{1}{M^2}\frac{\Box\phi}{f}+\frac{1}{M^4}\frac{\Box^2\phi}{f} - 2 \frac{1}{M^6}\frac{\partial^\lambda \partial^\alpha \Box\phi}{f}k_\lambda k_\alpha\\\nn
& + \frac{1}{M^4}\(\frac{1}{f} \Box\phi\frac{(\partial\phi)^2}{f^2}-\frac{1}{2}\frac{1}{f} \Box^2 \phi+\frac{2}{f}\partial^\lambda \partial^\alpha \Box\phi k_\alpha k _\lambda\)- \frac{1}{M^6}\frac{2}{3} \(\frac{1}{f}\partial^\kappa \partial^\lambda \Box\phi k_\lambda k_\kappa\)\\\nn
 &   -\frac{1}{3!M^6}\big[\gamma^\mu, \gamma^{\nu}\big] \big[\gamma^\alpha, \gamma^{\beta}\big] \big[\gamma^\delta, \gamma^{\kappa}\big]   \frac{ \partial_\mu\phi}{f}
\( \frac{ \partial_\alpha\phi}{f} \frac{  \partial_\beta\partial_\delta\phi}{f}  k_\nu k_\kappa+  \partial_\nu\(\frac{ \partial_\alpha\phi\partial_\delta\phi}{f^2} \)k_\kappa k_\beta \)\\\nn
&+\frac{1}{M^6}\frac{1}{2}\frac{1}{f} \big[\gamma^\mu, \gamma^{\nu}\big] \big[\gamma^\alpha, \gamma^{\beta}\big]  \Box\phi  \frac{ \partial_\mu\phi}{f} \frac{ \partial_\alpha\phi}{f} k_\nu  k_\beta\\
 &  -\frac{1}{3!M^8}\big[\gamma^\mu, \gamma^{\nu}\big] \big[\gamma^\alpha, \gamma^{\beta}\big] \big[\gamma^\delta, \gamma^{\kappa}\big]  
\(2 \partial_\gamma\( \frac{ \partial_\mu\phi}{f}
\frac{ \partial_\alpha\phi\partial_\delta\phi}{f^2} \)k_\nu k_\kappa k_\beta k_{\gamma} \)+\ldots\Bigg].
\end{align}
Here the `$\ldots$' indicates terms which vanish after either tracing over the Dirac indices, integrating angles, or taking the limit $M\to\infty$.  Integrating over momentum $k$, we have
\begin{align}\nn
& \lim_{M\to \infty}{\rm tr}\langle x |\gamma_5 e^{\slashed{D}^2/M^2}|x\rangle \\\nn
 & =  \frac{i\pi^2}{\(2\pi\)^4} {\rm tr}\Bigg[  \lim_{M\to \infty} M^2\frac{\Box\phi}{f}+\frac{1}{6}\frac{\Box^2\phi}{f}  + \frac{1}{f} \Box\phi\frac{(\partial\phi)^2}{f^2} +\frac{1}{4}\frac{1}{f}\big[\gamma^\mu, \gamma^{\nu}\big] \big[\gamma^\alpha, \gamma^{\beta}\big] \Box\phi  \frac{ \partial_\mu\phi}{f} \frac{ \partial_\alpha\phi}{f}\delta_{\nu\beta}\\\nn
 &   -\frac{1}{3}\frac{1}{4}\big[\gamma^\mu, \gamma^{\nu}\big] \big[\gamma^\alpha, \gamma^{\beta}\big]  \big[\gamma^\delta, \gamma^{\kappa}\big]  \frac{ \partial_\mu\phi}{f}
\( \frac{ \partial_\alpha\phi}{f} \frac{  \partial_\beta\partial_\delta\phi}{f} \delta_{\nu\kappa}+  \partial_\nu\(\frac{ \partial_\alpha\phi\partial_\delta\phi}{f^2} \)\delta_{\kappa\beta} \)\\
 &  -\big[\gamma^\mu, \gamma^{\nu}\big] \big[\gamma^\alpha, \gamma^{\beta}\big]  \big[\gamma^\delta, \gamma^{\kappa}\big]  
 \partial_\gamma\( \frac{ \partial_\mu\phi}{f}\frac{ \partial_\alpha\phi\partial_\delta\phi}{f^2} \)\frac{1}{12}(\delta_{\nu\kappa}\delta_{\beta\gamma}+\delta_{\nu\beta}\delta_{\kappa\gamma}+\delta_{\nu\gamma}\delta_{\kappa\beta}) +\ldots \Bigg].
\end{align}
Tracing over the Dirac indices, after some algebra, we get
 \begin{align}\nn
 \lim_{M\to \infty}{\rm tr}\langle x |\gamma_5 e^{\slashed{D}^2/M^2}|x\rangle  = & 4\frac{i\pi^2}{\(2\pi\)^4} \Bigg[  \lim_{M\to \infty} M^2\frac{\Box\phi}{f} +\frac{1}{6}\frac{1}{f} \Box^2 \phi    -\frac{2}{3} \partial^\mu\(\frac{ \partial_\mu\phi}{f} \( \frac{ \partial\phi}{f}\)^2\) \Bigg],
\end{align}
and thus the anomaly function is
 \begin{align}
\mathscr{A} = & - \frac{1}{2\pi^2} \Bigg[  \lim_{M\to \infty} M^2\frac{\Box\phi}{f} +\frac{1}{6}\frac{1}{f} \Box^2 \phi    -\frac{2}{3} \partial^\mu\(\frac{ \partial_\mu\phi}{f} \( \frac{ \partial\phi}{f}\)^2\) \Bigg].
\end{align}
Note that we recover the result we found perturbatively in Minkowski space.

%-------------------------------------------------------------
%-----Heat Kernel approach------------------------------------
%-------------------------------------------------------------

\subsection{Heat kernel approach}\label{sec:pathheat}

The second approach we use is to recast the sum in eq.\ \eqref{eq:reg} as the solution of a heat equation (see Ref.\ \cite{Vassilevich:2003xt}, and references within).  The advantage of this powerful method is that it allows us to include both the effects of gravitation, as well as gauge fields in the transformation of the path integral measure. 

Since $\bar{\psi}_n(x)\gamma_5{\psi}_n(x)={\rm Tr}[\bar{\psi}_n(x)\gamma_5{\psi}_n(x)]$, where the trace is over the spinor indices we can write \cite{Vassilevich:2003xt}
\Beq
\sum_n\bar{\psi}_n(x)\gamma_5{\psi}_n(x)=\lim_{\tau\rightarrow0}{\rm tr}[\gamma_5K(\tau;x,x)]\,.
\Eeq
The heat kernel function, defined as
\Beq
\label{eq:HeatKernel}
K_{\alpha\beta}(\tau;x,x')=e^{-\tau\slashed{D}^2_{x}}\sum_n{\psi}_{n\alpha}(x)\bar{\psi}_{n\beta}(x'),
\Eeq
is the formal solution to the heat equation
\Beq
\label{eq:HeatEq}
-\frac{\partial}{\partial\tau}K_{\alpha\beta}(\tau;x,x')=(\slashed{D}^2_{x})_{\alpha\rho}K_{\rho\beta}(\tau;x,x')\,,
\Eeq
where $\alpha$, $\beta$ and $\rho$ are spinor indices. The heat equation operator can be written in the form
\Beq\label{eqn:heatop}
\slashed{D}^2=-(\tilde{\nabla}_\mu\tilde{\nabla}^\mu+Q(x))\,,
\Eeq
where $\tilde{\nabla}_\mu$ is a matrix covariant derivative operator and $Q(x)$ is a matrix which does not contain any derivative operators (for our fermion, we define these operators in eq. \eqref{eq:DefOperators}). 

We are interested in the heat kernel near $\tau = 0$, and we formally expand in powers of $\tau$
\Beq
\label{eq:tauexpansion}
K(\tau;x,x)=\frac{1}{(4\pi\tau)^2}\sum_{k=0}^{\infty}\tau^k E_k(x)\,.
\Eeq
Using this expansion, we can evaluate
\Beq \label{eq:heattrace}
\mathscr{A}(x)=\sum_n\bar{\psi}_n(x)\gamma_5{\psi}_n(x)=\lim_{\tau\rightarrow0}\frac{1}{(4\pi\tau)^2}\sum_{k=0}^{2}{\rm tr}[\gamma_5\tau^k E_k(x)]\,.
\Eeq
For a heat operator of the general form in eq.\ \eqref{eqn:heatop}, the $E_k(x)$ coefficients in the $\tau$-expansion in eq.\ \eqref{eq:tauexpansion} are known. In particular
\Beq\label{eq:HKcoeff2}
E_0 & = 1\,,   \qquad 
E_1=Q+\frac{R}{6}\,,\\
E_2 & =\frac{\Box Q}{6}+\frac{RQ}{6}+\frac{Q^2}{2}+\frac{\Box R}{30}+\frac{R^2}{72}-\frac{R_{\mu\nu}R^{\mu\nu}}{180}+\frac{R_{\mu\nu\rho\sigma}R^{\mu\nu\rho\sigma}}{180}+\frac{W^{\mu\nu}W_{\mu\nu}}{12}\,,
\Eeq
where  $W_{\mu\nu}=[\tilde{\nabla}_\mu,\tilde{\nabla}_\nu]$. For the theory at hand, we have \cite{Vassilevich:2003xt}
\Beq\label{eq:DefOperators}
\tilde{\nabla}_\mu&=\partial_\mu+\frac{1}{8}[\gamma_b,\gamma_c]\sigma_\mu^{bc}+A_\mu+\frac{i}{2}[\gamma_b,\gamma_c]e^b_\mu e^c_\nu\gamma_5\frac{\partial^\nu\phi}{f}\,,\\
Q&=-\frac{R}{4}+\frac{1}{4}[\gamma_b,\gamma_c]e^b_\mu e^c_\nu F^{\mu\nu}+i\gamma_5\frac{(\partial_\mu\phi)^{;\mu}}{f}-2\left(\frac{\partial\phi}{f}\right)^2\,,\\
W_{\mu\nu}&=F_{\mu\nu}-\frac{1}{4}\gamma_b\gamma_c e^b_\alpha e^c_\beta R^{\alpha\beta}{}_{\mu\nu}\\
&-i\gamma_5\gamma^a e_a^\rho\left(\gamma_\nu\frac{(\partial_\rho\phi)_{;\mu}}{f}-\gamma_\mu\frac{(\partial_\rho\phi)_{;\nu}}{f}\right)-\gamma^a e_a^\rho\frac{\partial_\rho\phi}{f}\left(\gamma_\mu \gamma^b e_b^\sigma\frac{\partial_\sigma\phi}{f}\gamma_\nu-\gamma_\nu \gamma^b e_b^\sigma\frac{\partial_\sigma\phi}{f}\gamma_\mu\right)\,.
\Eeq
We need the traces
\Beq
{\rm tr}[\gamma_5Q]&=4i\frac{\Box \phi}{f}\,,\quad 
{\rm tr}[\gamma_5Q^2] =\epsilon^{\mu\nu\alpha\beta}F_{\mu\nu}F_{\alpha\beta}-16i\left(\frac{\partial\phi}{f}\right)^2\frac{\Box \phi}{f}+2iR\frac{\Box \phi}{f}\,,\\
{\rm tr}[\gamma_5W^{\mu\nu}W_{\mu\nu}]&=\frac{1}{4}\epsilon^{\mu\nu\alpha\beta}R^{\rho\sigma}{}_{\mu\nu}R_{\rho\sigma\alpha\beta}+i{\rm tr}[\gamma^\alpha\gamma^\beta\gamma^\rho\gamma^\nu]R_{\alpha\beta\mu\nu}\frac{(\partial_\rho\phi)_{;\mu}}{f}\\
&-4i{\rm tr}[\gamma^\rho\gamma^\nu]F_{\mu\nu}\frac{(\partial_\rho\phi)_{;\mu}}{f}+64i\left(\frac{\partial\phi}{f}\right)^2\frac{\Box \phi}{f}-64i\frac{(\partial_\nu\phi)_{;\mu}}{f}\frac{(\partial^\mu\phi)}{f}\frac{(\partial^\nu\phi)}{f}\,.
\Eeq
Thus, we can compute the traces over the $E_k$ in eq.\ \eqref{eq:heattrace}
\Beq
{\rm tr}[\gamma_5E_0]&=0, \quad 
{\rm tr}[\gamma_5E_1]=4i\frac{\Box \phi}{f}\\
{\rm tr}[\gamma_5E_2]&=2i\frac{\Box^2 \phi}{3f}+2i\frac{G^{\mu\nu}(\partial_\mu\phi)_{;\nu}}{3f}+\frac{1}{2}\epsilon^{\mu\nu\alpha\beta}F_{\mu\nu}F_{\alpha\beta}+\frac{1}{48}\epsilon^{\mu\nu\alpha\beta}R^{\rho\sigma}{}_{\mu\nu}R_{\rho\sigma\alpha\beta}\\
&-\frac{8}{3}i\left(\frac{\partial^\mu\phi}{f}\frac{\partial_\nu\phi}{f}\frac{\partial^\nu\phi}{f}\right)_{;\mu}\,.
\Eeq
After putting everything together, we find the Minkowski space anomaly
\Beq\label{eqn:heatanomy}
\langle\nabla_\mu j^{\mu}_5(x)\rangle= \,&-\frac{1}{\tau}\left(\frac{\Box \phi}{2\pi^2f}\right)-\frac{\Box^2 \phi}{12\pi^2f}-\frac{\nabla_\mu (G^{\mu\nu}\partial_\nu\phi)}{12\pi^2f}+\frac{1}{3\pi^2}\nabla_\mu \left(\frac{\partial^\mu\phi}{f}\frac{\partial_\nu\phi}{f}\frac{\partial^\nu\phi}{f}\right) \\
&+\frac{1}{384\pi^2}\epsilon^{\mu\nu\alpha\beta}R^{\rho\sigma}{}_{\mu\nu}R_{\rho\sigma\alpha\beta}-\frac{1}{16\pi^2}\epsilon^{\mu\nu\alpha\beta}F_{\mu\nu}F_{\alpha\beta} \,.
\Eeq
We have a linearly divergent piece $\propto \tau^{-1}$ (which renormalizes the kinetic part of equation of motion for the axion, $\Box\phi + V'=-\langle\nabla_\mu  j^{\mu}_{5}(x)\rangle/f$, analogously to the terms dependent on the Pauli-Villars mass in section \ref{sec:minkowski}, or the mass scale above in section \ref{sec:pathmom}) and finite anomalous terms. Note that all of the terms are total derivatives. 

\subsection{Jacobian terms in the Lagrangian}

The results in this section indicate that when changing from the fermion basis in eq.\ \eqref{eqn:fermaxaction} to the basis in eq.\ \eqref{eqn:fermaxaction2} the transformation properties of the path integral measure imply that the action should be supplemented with the terms
\begin{align}\nn\label{eqn:jacterms}
\Delta \mathcal{L}_{\rm jac} = &\frac{1}{12\pi^2}\left[\frac{1}{2}G^{\mu\nu}\frac{\partial_\mu\phi}{f}\frac{\partial_\nu\phi}{f}-\left(\frac{\partial\phi}{f}\right)^4-\frac{1}{2}\left(\frac{\Box \phi}{f}\right)^2\right]\\
&-\frac{\phi}{384\pi^2 f}\epsilon^{\mu\nu\alpha\beta}R^{\rho\sigma}{}_{\mu\nu}R_{\rho\sigma\alpha\beta}+\frac{\phi}{16\pi^2 f}\epsilon^{\mu\nu\alpha\beta}F_{\mu\nu}F_{\alpha\beta} .
\end{align}
Note that these terms have precisely the right form to cancel the contributions from the anomalous axial current and recover the correct massless limit.

Our results in this section depended explicitly on the inclusion of the axion in regularizing the path integral. It is not immediately clear that the inclusion of this term is justified. However, clearly we recover exactly the right terms to cancel the anomalous behavior of the Feynman graphs in the previous sections. We note that if the chiral symmetry were promoted to a local gauge symmetry, the axion interaction is related to the pure gauge part of an axial gauge field. In addition to the usual anomalous gauge invariant terms, the axion terms above correspond to pure-gauge terms. The axion in this case then appears to behave as a Stueckelberg-like field for the axial symmetry. We leave further exploration of axial gauge theories to future work.

%-------------------------------------------------------------
%-----Backreaction and Renormalization-------------------
%-------------------------------------------------------------

\section{Backreaction and equivalence of the formulations}

\label{sec:Counterterms}

%\kalo{This section has been rewritten}

In this section we show that the $\psi$ and $\chi$ models, eqs.\ \eqref{eqn:fermaxaction} and \eqref{eqn:fermaxaction2}, respectively, are different formulations of the same theory, provided we account for the Jacobian terms coming from the non-invariance of the fermion path integral measure under chiral rotation, eq. \eqref{eqn:axialrot}, relating the two. For completeness, we also discuss the backreaction on the gravitational background.

To compute the backreaction effects of the fermions on the classical $\phi$ and $g_{\mu\nu}$ background fields we adopt a semiclassical approach. We derive the classical equations of motion for $\phi$ and $g_{\mu\nu}$ by extremising the total action
\begin{align}\label{eqn:fulladreg}
S_{\rm total}[\phi,g]=S_g^{\rm bare}[g]+S_\phi[\phi,g]+W_{\rm int}[\phi,g].
\end{align}
In this expression, the axion action is
\begin{align}\label{eq:AxionBare}
S_\phi[\phi,g]&=\int d^4x\sqrt{-g}\left[\frac{(\partial\phi)^2}{2}-V(\phi)\right]\,,
\end{align}
and the gravitational action is given by \cite{Birrell:1982ix,Parker:2009uva}
\Beq\label{eq:EinstBare}
S_g^{\rm bare}[g]=  \int d^4x\sqrt{-g}\Bigg[\frac{M_{{\rm Pl},B}^2}{2}\big(R&+2\Lambda_B\big)+d_{1,B}R^2\\
&+d_{2,B}R^{\mu\nu}R_{\mu\nu}+d_{3,B}\Box R+d_{4,B}R^{\mu\nu\alpha\beta}R_{\mu\nu\alpha\beta}\Bigg]\,.
\Eeq
The gravitational action contains the bare dimensionful and dimensionless $M_{{\rm Pl},B}$, $\Lambda_B$, $d_{i,B}$ constants. These constants are not observable, since they receive finite or infinite corrections from the interaction effective action, $W_{\rm int}[\phi,g]$, defined as
\begin{align}\label{eqn:fulladreg}
e^{iW_{\rm int}[\phi,g]}\equiv\int d\bar{\psi}d\psi e^{iS_f[\bar{\psi},\psi,\phi,g]}\,.
\end{align}
According to eq. \eqref{eqn:fermaxaction} in the massless limit $\psi$ decouples from the axion, since the interaction effective action $W_{\rm int}$ becomes a function only of the metric $g_{\mu\nu}$,
\begin{align}\label{eqn:limWint}
\frac{\delta S_\phi}{\delta \phi}=-\frac{\delta}{\delta\phi}W_{\rm int}[\phi,g]=-\left\langle \frac{\delta S_f[\bar{\psi},\psi,\phi,g]}{\delta \phi}\right\rangle\underset{m\to 0}{=}0\,.
\end{align}
$\psi$ remains coupled to the gravitational background. The Einstein equations are sourced by $W_{\rm int}$,
\begin{align}
\label{eq:EinstBack}
\frac{2}{\sqrt{-g}}\frac{\delta S_g^{\rm bare}}{\delta g_{\mu\nu}}+\frac{2}{\sqrt{-g}}\frac{\delta S_\phi}{\delta g_{\mu\nu}}=-\frac{2}{\sqrt{-g}}\frac{\delta W_{\rm int}}{\delta g_{\mu\nu}}=-\left\langle \frac{2}{\sqrt{-g}}\frac{\delta S_f[\bar{\psi},\psi,\phi,g]}{\delta g_{\mu\nu}}\right\rangle\equiv\langle T^{\mu\nu}_\psi\rangle\,.
\end{align}
Nevertheless, the right hand side of eq.~\eqref{eq:EinstBack} can be absorbed into the gravitational terms on the left after renormalizing the gravitational action, $S_g^{\rm ren}=S_g^{\rm bare}-i\log \mathcal{Z}_f$, where $\mathcal{Z}_f[g]\equiv\lim_{m\to0}\int d\bar{\psi}d\psi e^{iS_f[\bar{\psi},\psi,\phi,g]}$. The renormalization due to the massless $\psi$ amounts to constant shifts in the coupling constants in eq.~\eqref{eq:EinstBare}, e.g., $d_{i,B}\to d_{i,B}+\Delta d_{i}$, etc., where explicit expressions for the constant shifts can be found in, e.g., Refs.~\cite{Birrell:1982ix,Parker:2009uva,Vassilevich:2003xt}.

We now show that the vanishing of the backreaction on the axion background, eq.~\eqref{eqn:limWint}, remains valid in the $\chi=e^{i\gamma_5\phi/f}\psi$ formulation of the theory, eq.~\eqref{eqn:fermaxaction2}, provided we account for the Jacobian factor coming from the non-invariance of the fermion path integral measure, i.e.,
\begin{align}\label{eqn:fulladreg}
e^{iW_{\rm int}[\phi,g]}=e^{iW_{\rm Jac}[\phi,g]}\int d\bar{\chi}d\chi e^{iS_f[\bar{\chi},\chi,\phi,g]}\,.
\end{align}
Since when $m\to 0$
\begin{align}
\frac{\delta}{\delta \phi} W_{\rm Jac}[\phi,g]&=\frac{2}{f}\mathscr{A}(\phi,g)=\frac{1}{f}\langle\nabla_\mu j^{\mu}_5(\bar{\chi},\chi)\rangle=-\left\langle \frac{\delta S_f[\bar{\chi},\chi,\phi,g]}{\delta \phi}\right\rangle\,,
\end{align}
where
\begin{align}\label{eqn:WJac}
W_{\rm Jac}[\phi,g]&=\int d^4x\sqrt{-g}\frac{1}{12\pi^2}\left[\frac{1}{2}G^{\mu\nu}\frac{\partial_\mu\phi}{f}\frac{\partial_\nu\phi}{f}-\left(\frac{\partial\phi}{f}\right)^4-\frac{1}{2}\left(\frac{\Box \phi}{f}\right)^2\right]\,,
\end{align}
we recover the no backreaction limit in the derivative-coupling formulation of the theory
\begin{align}\label{eqn:limWintchi}
\frac{\delta S_\phi}{\delta \phi}=-\frac{\delta}{\delta\phi}W_{\rm int}[\phi,g]=-\frac{\delta}{\delta\phi}W_{\rm Jac}[\phi,g]-\left\langle \frac{\delta S_f[\bar{\chi},\chi,\phi,g]}{\delta \phi}\right\rangle\underset{m\to 0}{=}0\,.
\end{align}
Hence, even though $\langle\nabla_\mu j^{\mu}_5(\bar{\chi},\chi)\rangle$ is anomalous, its effect on correlation functions of the axion, $\phi$, is canceled by the Jacobian terms, yielding a zero net backreaction on the axion background, in agreement with the $\psi$ formulation of the theory.

We note that, since $W_{\rm int}[\phi,g]$ is the same in the $\chi$ and $\psi$ formulations, when the axion is considered as a dynamical quantum field, its correlation functions are independent of the choice of the fermion basis. The partition function of the quantum axion
\begin{align}
\mathcal{Z}_\phi=\int\mathcal{D}\phi e^{iS_\phi[\phi,g]+iW_{\rm int}[\phi,g]+i\int d^4x \mathcal{K}(x)\phi(x)},   
\end{align}
where $\mathcal{K}(x)$ is a classical source, can be used in the standard manner \cite{Birrell:1982ix} to determine the connected, time-ordered Green functions of the axion
\begin{align}
i^j\langle0|T(\phi(x_1)...\phi(x_j))|0\rangle_c=\left(\frac{\delta^j\ln\mathcal{Z_\phi}}{\delta\mathcal{K}(x_1)...\delta\mathcal{K}(x_j)}\right)_{\mathcal{K}=0},   
\end{align}
independently of the choice of fermion basis.

%We note that the $\psi$ and $\chi$ models, eqs. \eqref{eqn:fermaxaction} and \eqref{eqn:fermaxaction2}, respectively, remain different formulations of the same theory for arbitrary fermion masses, as long as the relevant Jacobian factors are accounted for. The main difference between the massless and massive cases, is that when $m\neq0$ there is an explicit coupling between $\psi$ and $\phi$ and thus a non-vanishing fermion backreaction on the axion background. This corresponds to a finite difference (as opposed to the exact cancelation in the massless case) between the Jacobian terms and the $\langle\nabla_\mu j^{\mu}_5(\bar{\chi},\chi)\rangle$ anomalous contributions to the effective axion action in the $\chi$ formulation. \peter{I am not sure I agree with this statement:} The effective background theory in the $m\neq0$ cases should be of the form $S_g^{\rm ren}+S_\phi+\mathcal{W}_{\rm Jac}[\phi,g]$, where $\mathcal{W}_{\rm Jac}[\phi,g]$ has terms of the same form given in eq.~\eqref{eqn:WJac}, but with different numerical prefactors. 

One benefit of working in the derivatively coupled basis, eq.\ \eqref{eqn:fermaxaction2}, is that it allows us to read off the form of the effective action obtained by integrating out a very massive fermion. In the limit where the fermion $\chi$ is very massive, $m\to\infty$, the effective theory tends to $S_g^{\rm ren}+S_\phi+{W}_{\rm Jac}[\phi,g]$. In this limit, the physical and regulator fields cancel in their contributions to the diagrams above in sections \ref{sec:minkowski} and \ref{sec:pertgrav}, leaving behind the terms that arise from the Jacobian factor from the change of basis derived in section \ref{sec:pathint}. This is analogous to the situation that occurs, for example, for axion gauge-field interactions \cite{Quevillon:2019zrd, Quevillon:2021sfz}.    

The effective axion action that results from integrating out the fermion in eq.\ \eqref{eqn:fermaxaction} or \eqref{eqn:fermaxaction2} contains a number of terms that have been invoked in various phenomenological contexts. Non-minimal theories of a scalar field involving only the first term in eq.~\eqref{eqn:WJac} have been considered in the context of inflation, dark matter and dark energy \cite{Sushkov:2009hk,Saridakis:2010mf,Gao:2010vr,Germani:2010gm,Germani:2010ux,Germani:2011ua,Germani:2011mx,Folkerts:2013tua}. The QCD axion non-minimally coupled to gravity with an interaction term of the form $\sim cG^{\mu\nu}\partial_\mu\phi\partial_\nu\phi/(24\pi^2f^2)$, can be a viable dark matter candidate provided the dimensionless coefficient is unnaturally large, $|c|\gg 1$, as required by the CMB isocurvature constraints \cite{Folkerts:2013tua}. Small-scale inflation can also be realized by a scalar field with a large non-minimal coupling \cite{Folkerts:2013tua}. Finally, kinetically driven inflationary scenarios have been constructed using the second and third terms in eq.~\eqref{eqn:WJac} \cite{Watanabe:2020ctz}.

%----------------------------------
%-----Summary and Conclusions------
%----------------------------------

\section{Summary and conclusions}\label{sec:conclusions}

In this paper, we have studied the theory of a Dirac fermion interacting with an axion-like particle, or a pseudoscalar via a derivative coupling. We have demonstrated that, in the massless limit, the classical chiral symmetry of the action  is spoiled by the axion background. We have computed the contributions to this anomalous Ward identity perturbatively using Feynman graphs in both Minkowski spacetime, as well as the linear gravitational corrections due to metric fluctuations around a Minkowski background. Using Fujikawa's method, we have also studied the transformation properties of the path integral measure under chiral rotations of the fermion in the presence of the axion background, and shown that the resulting anomaly function matches the result computed with Feynman graphs. 

Our results indicate that the axial current of  fermions derivatively coupled to axions is naively anomalous. In particular, our results show that the two theories (in eqs.\ \eqref{eqn:fermaxaction} and \eqref{eqn:fermaxaction2}) are two formulations of the same underlying theory, provided that one accounts for the Jacobian terms coming from the non-invariance of the fermion path integral measure under chiral rotations.  The form of the Jacobian from the field redefinition also indicates that the  one-loop effective action of a pseudo-scalar should naturally contain the dimension-six operators, $(\Box\phi/f)^2$ and $G^{\mu\nu}\partial_\mu\phi\partial_\nu\phi/f^2$, as well as the dimension eight operator $(\partial\phi/f)^4$. In particular, these interactions are induced when a heavy fermion is integrated out of the theory. While such interactions are likely irrelevant for axion-fermion theories like the Peccei-Quinn solution to the strong CP problem \cite{PhysRevLett.38.1440}, these terms may lead to interesting effects in theories of axion inflation or axion dark matter.

This result resolves difference in the results found in \cite{Adshead:2015kza}, and  \cite{Adshead:2018oaa, Adshead:2019aac}. Working in the derivatively coupled basis, Ref.\ \cite{Adshead:2015kza} found non-vanishing fermion backreaction effects during axion-driven inflation in the limit that the fermion was massless. On the other hand, Refs.\ \cite{Adshead:2018oaa, Adshead:2019aac} argued that one should make a field redefinition to make the chiral symmetry manifest in the limit $m\to 0$. However, away from the massless limits, the two formulations of apparently the same theory appeared to give different results. The results presented here indicate that the origin of these differences is in the transformation properties of the path integral measure.

Our results have been computed using two different regularization schemes for the Feynman graphs, t'Hooft-Veltman dimensional regularization and Pauli-Villars, and using the fermion modes in the derivatively coupled basis to regularize the path integral computation. The resulting anomaly function is a local function of field operators, which suggests that perhaps a regularization scheme exists which preserves the chiral symmetry and conserves this current. However, in the presence of gauge interactions, such a scheme would need to also preserve gauge invariance.  We leave investigation of this possibility to future work.

\acknowledgments

We thank Giovanni Cabass, Patrick Draper, Yoni Kahn, Rob Leigh, and Jessie Shelton for useful conversations; and Patrick Draper, Lauren Pearce and Lorenzo Sorbo  for comments on a draft of this manuscript. This work was supported in part by the US Department of Energy through grant DESC0015655.

\appendix

%-----------------------------------------------------------------
%-----Adiabatic solutions to the Dirac equation in de Sitter------
%-----------------------------------------------------------------

\section{Adiabatic solutions to the Dirac equation in de Sitter}\label{app:adiabatic}

In this appendix, we summarize the adiabatic expansion of the Dirac equation for the fermion interacting with the axion field originally presented in Ref.\ \cite{Adshead:2018oaa}, and list the parts of the solutions up to fourth order that are required for the computation of the anomaly in the text.

We begin with the Dirac equation for the fermion
\begin{align}
i \dot{ u}_r &= m  u_r + \left( \dfrac{k}{a} + 2 H r \xi \right)  v_r, \quad
i \dot{ v}_r = - m   v_r + \left( \dfrac{k}{a} + 2 H r \xi \right)   u_r.
\end{align}
Our goal is to solve these equations as an expansion in the Hubble parameter $H$.  We introduce the ansatz \cite{Landete:2013axa, Landete:2013lpa, delRio:2014cha}
\begin{align}
 u_r &= \sqrt{ 1 + \dfrac{m}{\omega}} \exp \left(
-i \int^t d\tilde{t} \left[ \omega(\tilde{t}) + H \omega_1(\tilde{t}) + H^2 \omega_2 (\tilde{t}) \right] \right)
\left[ 1 + H F_{1,r} + H^2 F_{2,r} \right], \nonumber \\
 v_r &= \sqrt{ 1 - \dfrac{m}{\omega}} \exp \left(
-i \int^t d\tilde{t} \left[ \omega(\tilde{t}) + H \omega_1(\tilde{t}) + H^2 \omega_2 (\tilde{t}) \right] \right)
\left[ 1 + H G_{1,r} + H^2 G_{2,r} \right],
\end{align}
where we have the freedom to choose $F_{i,r}$ to be real, and where $\omega = \sqrt{ k^2/a^2 + m^2 }$.  We solve  the equations of motion iteratively, and impose the quantization condition
\begin{align}
 u_r  u_r^* +  v_r  v_r^* = 2,
\end{align}
at each order. In order to reproduce the anomaly, we require the solutions up to fourth order in the expansion in $H$. The solutions up to second order in $H$ were found in Ref.\  \cite{Adshead:2018oaa}, and read
\begin{align}
F_{1,r} &= - \dfrac{r \xi m}{\omega^2} \sqrt{ \dfrac{\omega - m}{\omega + m} }\,,\qquad G_{1,r} = \dfrac{r \xi m}{\omega^2} \sqrt{ \dfrac{\omega + m}{\omega - m} }  + i\dfrac{m}{2  \omega^2}, \nonumber \\
\omega_1 &= 2 r \xi \sqrt{ 1 - \dfrac{m^2}{\omega^2}}\,,\qquad \omega_2= \dfrac{m (4 \omega - 5 m) (\omega^2 - m^2)}{8 \omega^5} + \dfrac{2 m^2 \xi^2}{\omega^3}, \nonumber \\
F_2 &= \dfrac{m (\omega - m)(5 m^2 - 4 \omega^2)}{16 \omega^6} + \dfrac{m \xi^2 (4 \omega - 5 m)}{2 \omega^4}, \nonumber \\
G_2 &= \dfrac{m (-5 m^3 - 5 m^2 \omega + 2 m \omega^2 + 4 \omega^3)}{16 \omega^6}
-i \dfrac{m r \xi (5 m + 6 \omega) }{2 \omega^4} \sqrt{
\dfrac{ \omega - m}{\omega + m} } - \dfrac{m \xi^2 (5 m + 4 \omega)}{2 \omega^4} \,.
\end{align}
To get all contributions to the anomaly in eq.\ \eqref{eqn:adiabaticreg} above, we require the solutions up to order $H^4$. Given the long and not particularly enlightening nature of these solutions, we list only the required pieces for evaluating eq.\ \eqref{eqn:adiabaticreg}. These are
\begin{align}\nn
\omega^{(3)}  = &  -\frac{4 m^2 \xi ^3 r  \sqrt{\omega^2 - m^2}}{\omega ^5}+\\
&\frac{m \xi  r \left(25 m^5-16 m^4 \omega -50 m^3 \omega
   ^2+32 m^2 \omega ^3+27 m \omega ^4-18 \omega ^5\right)}{4 \omega ^7\sqrt{\omega^2 - m^2}}\\ \nn
F_{3,r} =  & -\frac{m \xi ^3 r \left(15 m^3-11 m^2 \omega -12 m \omega ^2+8
   \omega ^3\right)}{2 \omega ^6 \sqrt{\omega^2 - m^2}}\\\nn
   & -\frac{m \xi  r \left(65 m^5-60 m^4 \omega -111
   m^3 \omega ^2+102 m^2 \omega ^3+48 m \omega ^4-44 \omega ^5\right)}{16 \omega ^8
   \sqrt{\omega^2 - m^2}}\\\nn
\Re[G_{3,r}] = &  -\frac{m \xi ^3 r \left(15 m^3+11 m^2 \omega -12 m \omega ^2-8
   \omega ^3\right)}{2 \omega ^6 k(t)}\\\nn\ & -\frac{m \xi  r \left(65 m^5+60 m^4 \omega -93 m^3
   \omega ^2-100 m^2 \omega ^3+24 m \omega ^4+44 \omega ^5\right)}{16 \omega ^8
   \sqrt{\omega^2 - m^2}},\\\nn
\Im\[G_{3, r}\] = &  -\frac{m \xi ^2 \left(45 m^3+57 m^2 \omega -32 m \omega ^2-48
   \omega ^3\right)}{4 \omega ^6 (m+\omega )}\\ \nn & -\frac{m \left(65 m^4+5 m^3 \omega -86 m^2
   \omega ^2-4 m \omega ^3+24 \omega ^4\right)}{32 \omega ^8}.
\end{align}
At fourth adiabatic order, we need only the imaginary part of $G_{4,r}$ for eq.\ \eqref{eqn:adiabaticreg}
\begin{align}
\Im\[G_{4,r}\] = & \frac{m \xi ^3 r (m-\omega )^2 \left(195 m^4+449 m^3 \omega +170 m^2
   \omega ^2-240 m \omega ^3-160 \omega ^4\right)}{4 \omega ^8 k(t)^3}\\ \nn &+\frac{m \xi  r
   (m-\omega )^2}{32 \omega ^{10} k(t)^3} \Big(1105 m^6+2350 m^5 \omega -53 m^4 \omega ^2\\\nn 
   &-2826 m^3 \omega
   ^3-1220 m^2 \omega ^4+708 m \omega ^5+400 \omega ^6\Big).
 \end{align}

\bibliography{AxFermAnom}
\bibliographystyle{JHEP}

\end{document}